\documentclass[fleqn,usenatbib]{mnras}

\usepackage[T1]{fontenc}
\usepackage{ae,aecompl}

\usepackage{graphicx,color}	
\usepackage{amsmath}	
\usepackage{amssymb}	
\usepackage{soul}

\newcommand\mearth{{\,{\rm M}_{\oplus}}}
\newcommand\mj{{\,{\rm M}_{\rm J}}}

\newcommand\simlt{\la}

\title[Imaging protoplanets with ALMA]{Searching for wide-orbit gravitational instability protoplanets with ALMA in the dust continuum}

\author[]{J. Humphries$^1$, C. Hall$^{1,2,3}$ \thanks{cassandra.hall@uga.edu}, T. J. Haworth$^4$ \& S. Nayakshin$^1$
\\
\\
$^{1}$
Department of Physics and Astronomy, University of Leicester, Leicester LE1 7RH, UK. \\
$^{2}$Department of Physics and Astronomy, The University of Georgia, Athens, GA 30602, USA. \\
$^{3}$Center for Simulational Physics, The University of Georgia, Athens, GA 30602, USA.\\ 
$^{4}$ Astronomy Unit, School of Physics and Astronomy, Queen Mary University of London, London E1 4NS, UK
}

\date{Accepted XXX. Received YYY; in original form ZZZ}

\pubyear{2020}

\begin{document}
\label{firstpage}
\pagerange{\pageref{firstpage}--\pageref{lastpage}}
\maketitle

\begin{abstract}
Searches for young gas giant planets at wide separations have so far focused on techniques appropriate for compact (Jupiter sized) planets. Here we point out that protoplanets born through Gravitational Instability (GI) may remain in an initial pre-collapse phase for as long as the first $ 10^5-10^7$ years after formation. These objects are hundreds of times larger than Jupiter and their atmospheres are too cold ($T\sim$ tens of K) to emit in the NIR or H$\alpha$ via accretion shocks. However, it is possible that their dust emission can be detected with ALMA, even around Class I and II protoplanetary discs. 
In this paper we produce synthetic observations of these protoplanets. We find that making a detection in a disc at 140 parsecs would require a few hundred minutes of ALMA band 6 observation time. Protoplanets with masses of 3-5 $M_J$ have the highest chance of being detected; less massive objects require unreasonably long observation times (1000 minutes) while more massive ones collapse into giant planets before $10^5$ years.
We propose that high resolution surveys of young ($10^5-10^6$ years), massive and face on discs offer the best chance for observing protoplanets. Such a detection would help to place constraints on the protoplanet mass spectrum, explain the turnover in the occurrence frequency of gas giants with system metallicity and constrain the prevalence of GI as a planet formation mechanism. Consistent lack of detection would be evidence against GI as a common planet formation mechanism. 

\end{abstract}

\begin{keywords}
accretion discs -- planet-disc interactions -- protoplanetary discs -- brown dwarfs -- planets and satellites: formation -- planets and satellites: composition
\end{keywords}

\section{Introduction}

Huge leaps in imaging capabilities of protoplanetary discs have been made in the last decade. This has revealed a plethora of substructure, such as rings \citep{DipierroEtal15,DipierroEtal18,andrewsetal2018}, and spirals \citep{garufietal2013,benistyetal2015,PerezEtal2016}. Thanks to kinematic detections \citep{TeagueEtal18,PinteEtal18,pinteetal2019,PinteEtal20}, it is now largely understood that ring-like features are due to forming core-accretion protoplanets. The origin of spiral morphology is more ambiguous, and may be due to planets \citep[see e.g.][]{dongetal2015_planet,veronesietal2019}, or, alternatively,  gravitational instability \citep{donghallricechiang2015,halletal2016,MeruEtal17,halletal2018}.

In the Gravitational Instability (GI) theory of planet formation,
if a disc cools sufficiently rapidly relative to its dynamical timescale, it will undergo fragmentation \citep{gammie2001,riceetal2003}, which is more likely in discs around higher mass stars \citep{cadmanetal2020,haworthetal2020}. Fragmentation forms gravitationally bound objects 
with radii of around 1 AU, at distances beyond 50 au from the central star \citep{Rice05,Rafikov05,HelledEtalPP62014}.



If these protoplanets remain embedded in the disc they will rapidly migrate inwards in the type I regime \citep{BaruteauEtal11}. In this case, tidal disruption may destroy a large fraction of these objects \citep{NayakshinFletcher15, HumphriesEtal19}.

On the other hand, planet-planet interactions, gap-opening or early disc dispersal may leave many protoplanets beyond the outer disc edge at orbits of tens or even hundreds of AU \citep{VB06, BoleyDurisen10, HallEtal17, ForganEtal18}.
After a protoplanet is born it cools and slowly contracts until hydrogen in its core dissociates at $\sim$ 2000 K. At this point it undergoes a rapid collapse event to form a gas giant with a radius a few times larger than Jupiter. 
These two stages are known as `pre' and 'post' collapse, and represent a low mass equivalent to first and second core formation in the field of star formation \citep{Larson69, Nayakshin10a,BhandareEtal18}.

High resolution radiative transfer and convection simulations of this pre-collapse phase find that it may last as long as $10^6$ years for Jupiter mass protoplanets \citep{VazanHelled12}, significantly longer than for stellar first cores.  
Furthermore, GI protoplanets accrete pebbles very efficiently \citep{HumphriesNayakshin18, BaehrKlahr19} and so are expected to have a  higher metallicity than the surrounding disc. The associated increase in opacity may increase their cooling time \citep{HelledBodenheimer11}, with some authors recently finding that the pre-collapse phase may last as long as $10^7$ for metal enriched protoplanets less massive than $\sim 2\mj$ \citep{Nayakshin20-TW-Hya}.
There is therefore a possibility that these extended, pre-collapse, dusty protoplanets could be observable with the current generation of Atacama Large Millimeter/submillimeter Array (ALMA) surveys, offering a key test of the GI planet formation mechanism. 
The question is, how many should we expect to detect? 

While protoplanets represent a mid-point in the GI planet formation process from gravitationally unstable disc to gas giant, constraining their number from either end offers only a limited picture.
The giant planet population at wide orbits ($a>10 $AU) appears to be rare at late times, \cite{ViganEtal17} found that wide orbit gas giants appear in only 2 \% of systems, though \cite{NielsonEtal19} recently found a figure of 9 \% around high mass ($M_{*} > 1.5 M_{\odot}$) stars.
However, neither the spatial or mass distributions of the giant planet population should be expected to match those of the precursor protoplanet population.
For instance, \cite{GalvagniEtal12} showed that protoplanets may lose as much as half of their mass during collapse. Furthermore, not all protoplanets survive to reach collapse; processes that prolong the cooling phase such as stellar irradiation and core growth can lead to disruption events even millions of years after formation \citep{Nayakshin16a}. And of course, the fraction of the giant planet population born via the separate Core Accretion formation process also remains uncertain \citep{SantosEtal17,Schlaufman18,SuzukiEtal18}. 
Current surveys for compact, Jupiter-like giants are not suited to detecting pre-collapse protoplanets since they focus on post-collapse objects with $\sim 1 R_{\textrm J}$ and temperatures of 1000-2000 K. Depending on the evolutionary phase and dust opacity model, pre-collapse protoplanets have radii from a fraction to as much as a few AU and effective temperatures of 10-50 K. 

At the other end of the evolutionary scale, large ALMA programmes have observed only a handful of gravitationally unstable discs. The majority of systems in million year old star formation regions are low mass (dust masses $<$ 10 $M_{\oplus}$) and have outer dust disc radii of less than $\sim$ 15 AU \citep{AnsdellEtal17,AnsdellEtal18,LongEtal18b,CiezaEtal19,CazzolettiEtal19}. These discs are likely to have been more massive in their youth, since massive discs evolve rapidly \citep{HallEtal19}, and are frequently truncated by dynamical interactions during their first $10^4-10^5$ years of life \citep{Bate18}. After formation, discs are then truncated further by far-ultraviolet (FUV) radiation from external photoevaporation \citep{WinterEtal18}. Millimetre dust is also not an accurate tracer of the true outer disc edge due to inwards radial drift of dust particles and sharp opacity changes with grain size \citep{FacchiniEtal17,TrapmanEtal19,RosottiEtal19}. If the fraction of discs that were gravitationally unstable during their formation is uncertain, the initial population of protoplanets is even more so, fragmenting discs in simulations typically form anywhere from a few to a few dozen protoplanets \citep{VB10, BoleyDurisen10, ZhuEtal12a, HallEtal17}.

While the efficiency of GI as a planet formation mechanism remains uncertain, in this paper we suggest that making a direct detection of a protoplanet offers a unique opportunity to place constraints on this value.
In fact, there are already tantalising signs that point towards the presence of protoplanets in million year old protoplanetary discs.
The numerous annular gaps and bright rings detected in discs by ALMA are widely believed to be formed by gas giant planets with masses in the range $0.1- 10\mj$ \citep[][see \cite{NayakshinEtal19} for an estimate of the mass function of these planet candidates]{DipierroEtal15,DipierroEtal16a,ClarkeEtal18,LodatoEtal19}\footnote{Although non-planet based interpretations of these signatures are possible, they cannot account for most of the annular structures detected by ALMA \citep{DSHARP7}.}. Planets have now been directly detected in some of the best studied discs, e.g., PDS70 \citep{KepplerEtal19}, with masses and orbits in agreement with the dust modelling predictions.
Furthermore, the presence of planets in disc gaps has also been inferred from `velocity kinks' in CO line emission \citep{PerezEtal2016}. These results indicate that many of the gaps in the ALMA observed discs may host $\sim 1-3\mj$ mass planets \citep{CasassusPerez19,PinteEtal20}. This exciting progress unearths one significant puzzle, however. The frequency of gas giant planets required by the observations of young protoplanetary discs exceeds the frequency of detection of such planets via direct searches in the NIR/optical wavelengths by a factor of $\sim 10$ \citep{BrittainEtal20}. The authors of that study suggest an episodic accretion solution to this discrepancy. Another solution could be that these planets are in the extended pre-collapse phase and are simply not hot enough to be detected in the NIR/optical. An important question to ask is: while previous ALMA observations are clearly sufficient to characterise disc structure, could they directly detect dust emission of pre-collapse GI planets? What integration times are required in order to detect these planets?

\cite{ZakhozhatEtal13, DouglasEtal13, MayorEtal16} and \cite{HallEtal19} previously studied the observability of gravitationally unstable discs during instability and immediately after fragmentation, while \cite{MeyerEtal19} and \cite{JankovicEtal19} have recently done the same for fragmenting discs around high mass protostars. 
However, these studies were primarily concerned with detecting systems in the GI unstable phase ($\simlt 10^5$ years), no work has yet considered the observability of pre-collapse protoplanets at ages $10^5-10^7$ years.

In this study we quantify the resolution and integration time requirements for ALMA mm dust continuum observations necessary to detect pre-collapse protoplanets around protoplanetary discs at million year ages. Spotting these stranded protoplanets would help to constrain the occurrence frequency of GI as a planet formation mechanism, since Core Accretion (CA) planets never pass through this extended, first-core like stage. Observing such extended protoplanets would also demonstrate that GI is able to frequently form objects in the planetary mass regime, rather than being limited to Brown Dwarfs and stellar mass companions \citep{ForganRice13,KratterEtal10}.

In Section \ref{sec:methods} we outline our method for producing synthetic observations of protoplanets and study the impact of choosing different ALMA antenna configurations and integration times. In Section \ref{sec:results} we assess the most appropriate configuration choices based on the Signal to Noise Ratio (SNR) of our synthetic protoplanet population. Following this, in Section \ref{sec:obs} we consider how protoplanet observability varies as a function of system age by combining the results of our synthetic observations with protoplanet evolution models from \cite{Nayakshin15c}. Finally we discuss the relevance of these results to observations in the nearby Ophiuchus Molecular Cloud and outline our suggestions for maximising the chance of making a protoplanet detection with ALMA.

\section{Methods}
\label{sec:methods}

\subsection{Pre-collapse GI protoplanets}\label{sec:pre-collapse}

Although protoplanets form from extended patches of the outer disc, they rapidly contract from a size of $\sim$ 10 to $\sim 1$ AU on close to the dynamical timescale. At this point their evolution slows and their subsequent contraction to a radius of $\sim$0.1 AU may take up to $10^5-10^7$ years, depending on their mass and bulk metallic composition \citep{HelledEtal11,Nayakshin20-TW-Hya}. During this pre-collapse stage, their central temperatures rise steadily from a few hundred up to 2000 K (where collapse occurs due to Hydrogen dissociation) while their surface temperatures remain at a few tens of Kelvin. Their interiors are highly optically thick and dominated by convection, energy is only lost radiatively from a thin surface layer \citep{Bodenheimer74}. 
In this paper we consider two stages of pre-collapse evolution, a young and large ($R_{PP} = 1.5$ AU) protoplanet as well as a smaller, older ($R_{PP}=1$ AU) example.
We model the protoplanets as polytropes with an ideal equation of state and adiabatic index $\gamma = 7/5$, the appropriate choice for a molecular Hydrogen dominated, pre-collapse protoplanet formed via GI \citep{BoleyEtal07}. 

We position these protoplanets at a `near' (40 AU) and a `far' (100 AU) separation from the host star, representing protoplanets left stranded beyond the disc edge by planet-planet scattering during disc fragmentation \citep[c.f.,][]{HallEtal17,ForganEtal18}.
We make this choice since protoplanets beyond the disc edge are more likely to survive to million year ages, as the migration time for planets embedded within the disc is typically of the order $10^4$ years \citep[e.g.,][]{BoleyEtal10,BaruteauEtal11}. Even massive, gap-opening protoplanets may migrate significantly, their final position after disc dispersal depends sensitively on the poorly constrained disc viscosity parameter $\alpha$ \citep{FletcherEtal19,HumphriesEtal19}. Considering a protoplanet beyond the disc edge allows us to avoid these modelling uncertainties.
 
We set the surface temperatures of our protoplanets to 30 K at 40 AU and 20 K at 100 AU to account for surface heating from stellar irradiation. These temperatures are consistent with the temperature profile of our disc (see the following section) and in any case are also typical of the protoplanet surface temperatures calculated in isolated 1D simulations without irradiation \citep{VazanHelled12}.

\subsection{The dust disc}

\begin{figure}
\includegraphics[width=1.0\columnwidth]{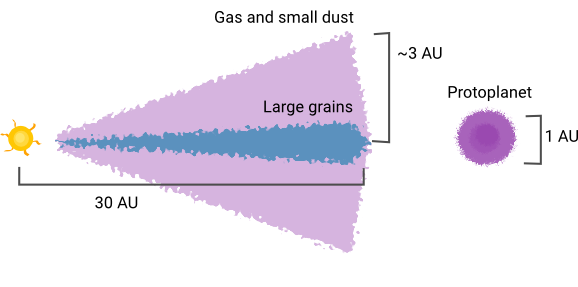}
\caption{Graphical representation of the system setup used in this paper. In the disc, small dust is distributed according to the gas density whilst large grains are suppressed in the vertical direction by a factor of ten in order to represent settling. All grains are distributed following the gas density inside the protoplanet to account for internal convection.}
\label{fig:setup}
\end{figure}

\begin{figure}
\includegraphics[width=1.1\columnwidth]{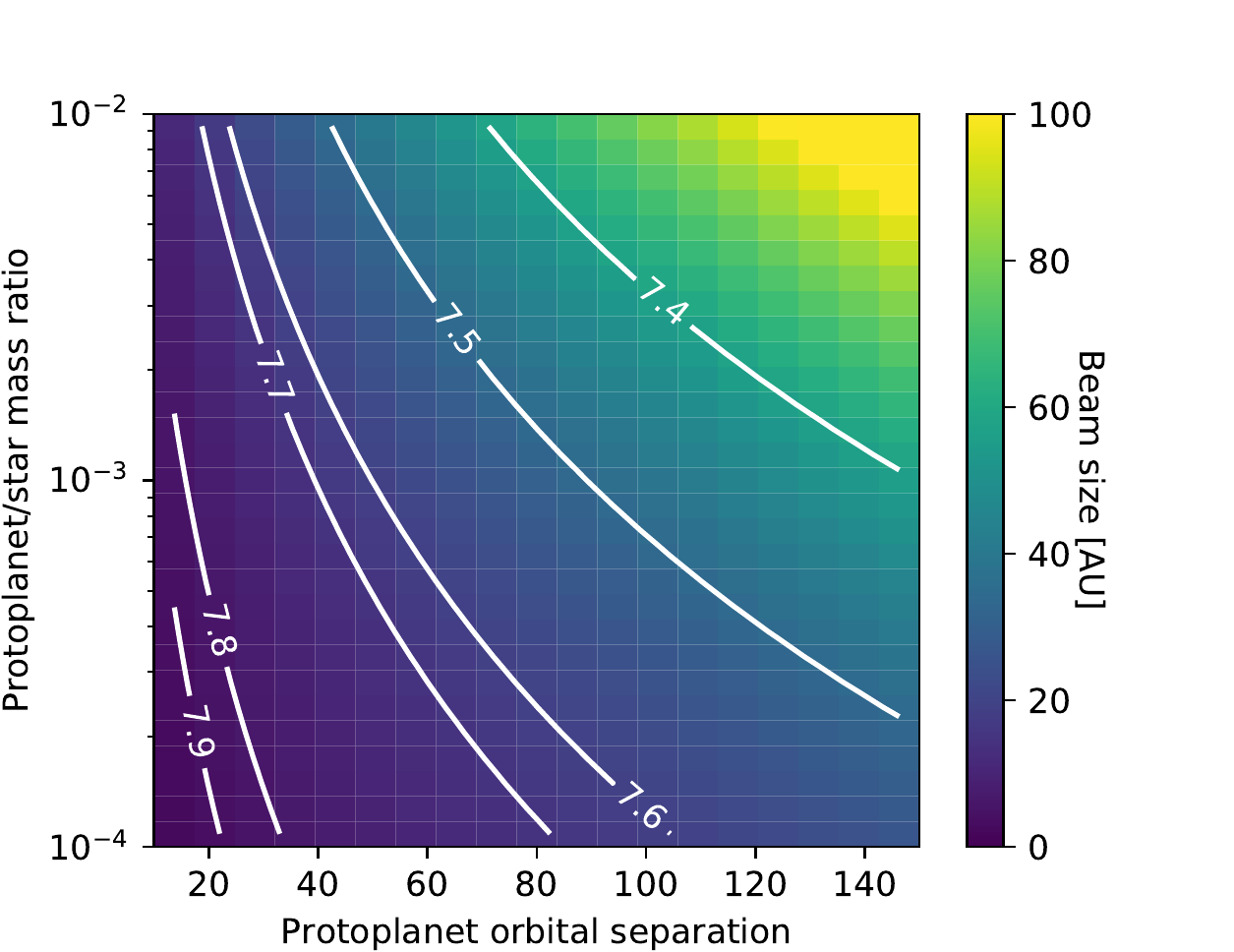}
\caption{Lower limits on the beam size (AU) required to resolve a protoplanet orbiting around a disc. Beam sizes for systems at 140 pc are marked as white contours for different ALMA cycle 7 antenna configurations, systems above each contour line are resolvable for the specified configuration. This figure assumes that the protoplanet is close to the disc edge and that the corresponding disc truncation follows the relation found empirically in \protect\cite{LodatoEtal19}. As such this figure represents a lower limit for the required beam size, for Cycle 7, configuration C7.}
\label{fig:PP_lengthscale}
\end{figure}

\begin{figure*}
\begin{tabular}{ccc}
    \includegraphics[width=0.71\columnwidth]{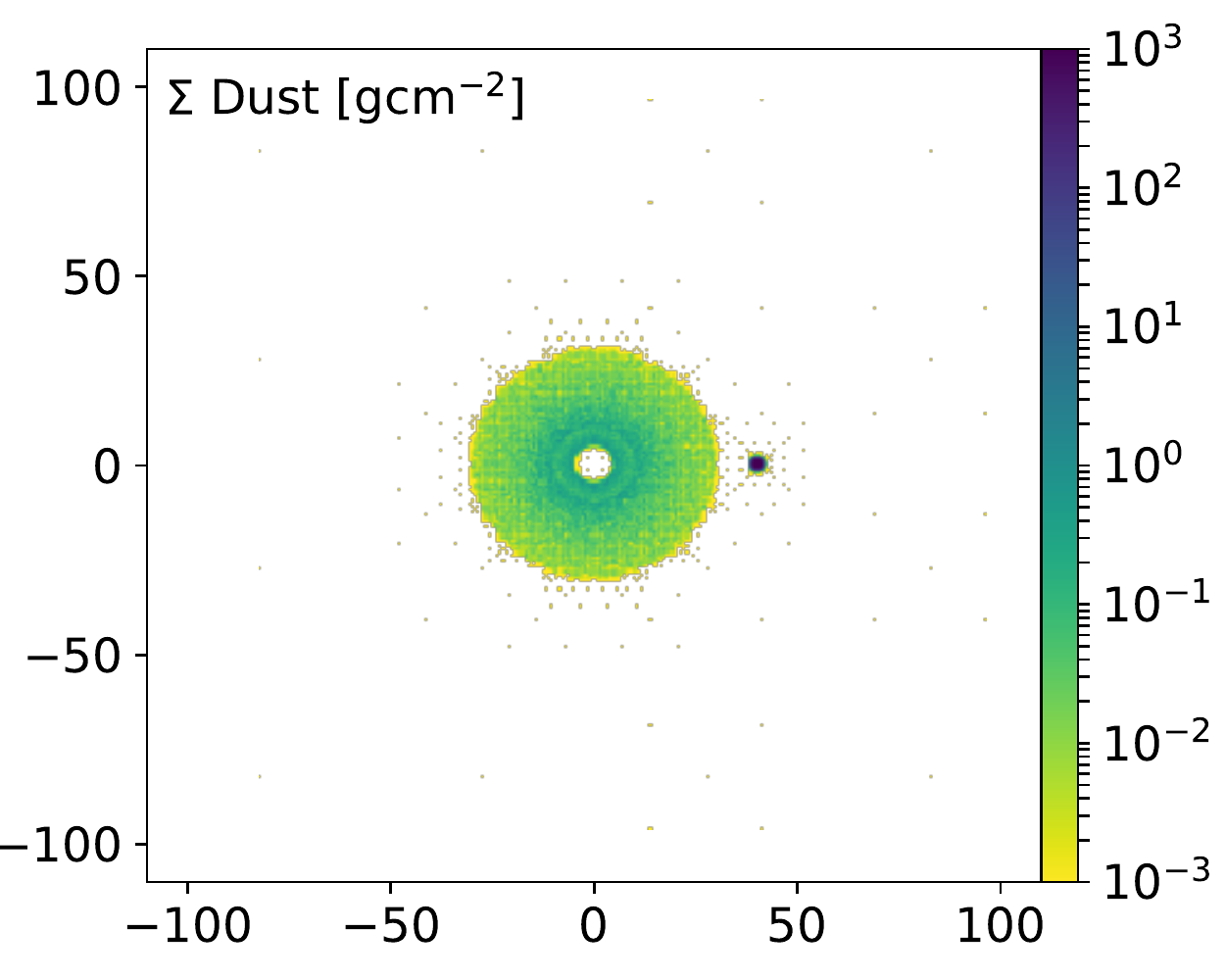} &
    \includegraphics[width=0.64\columnwidth]{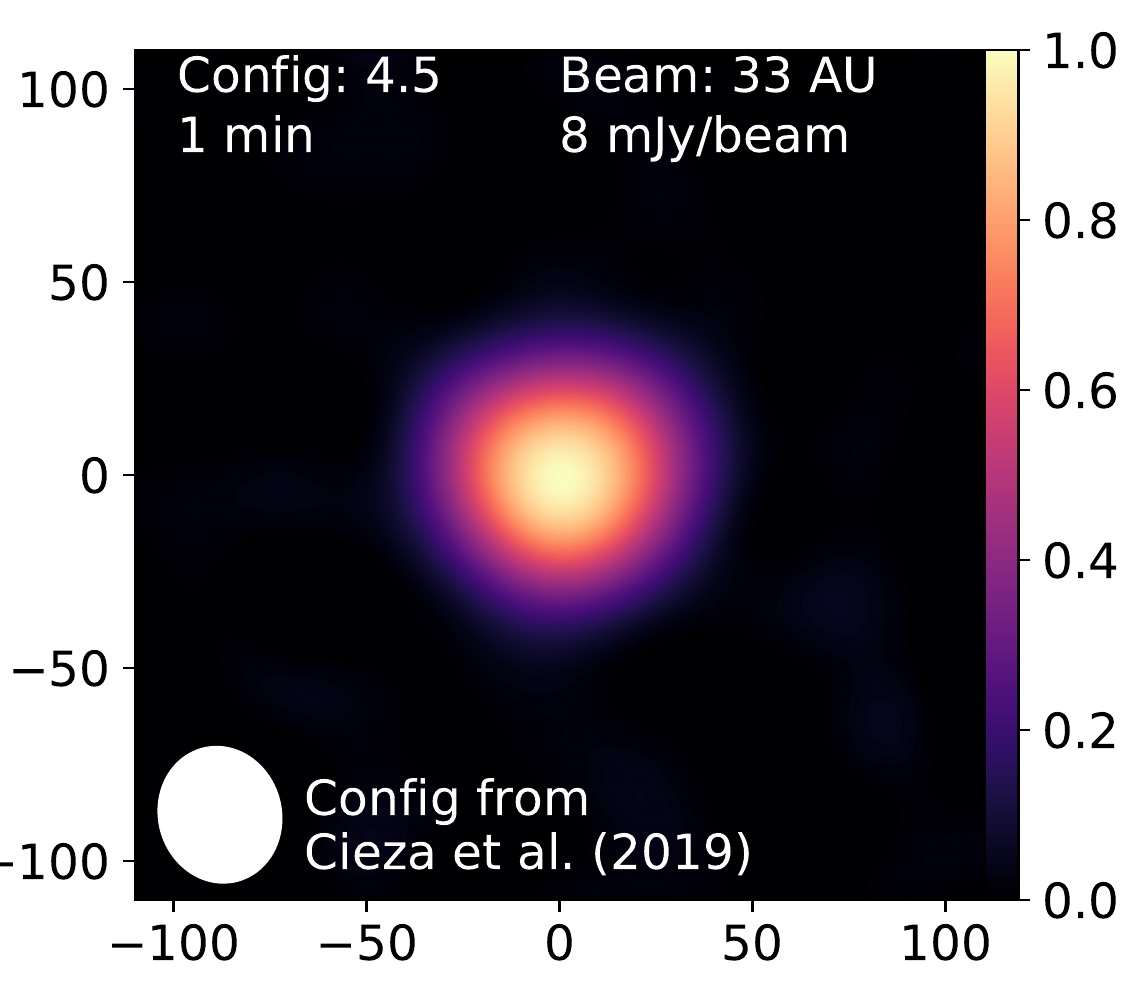} &
    \includegraphics[width=0.64\columnwidth]{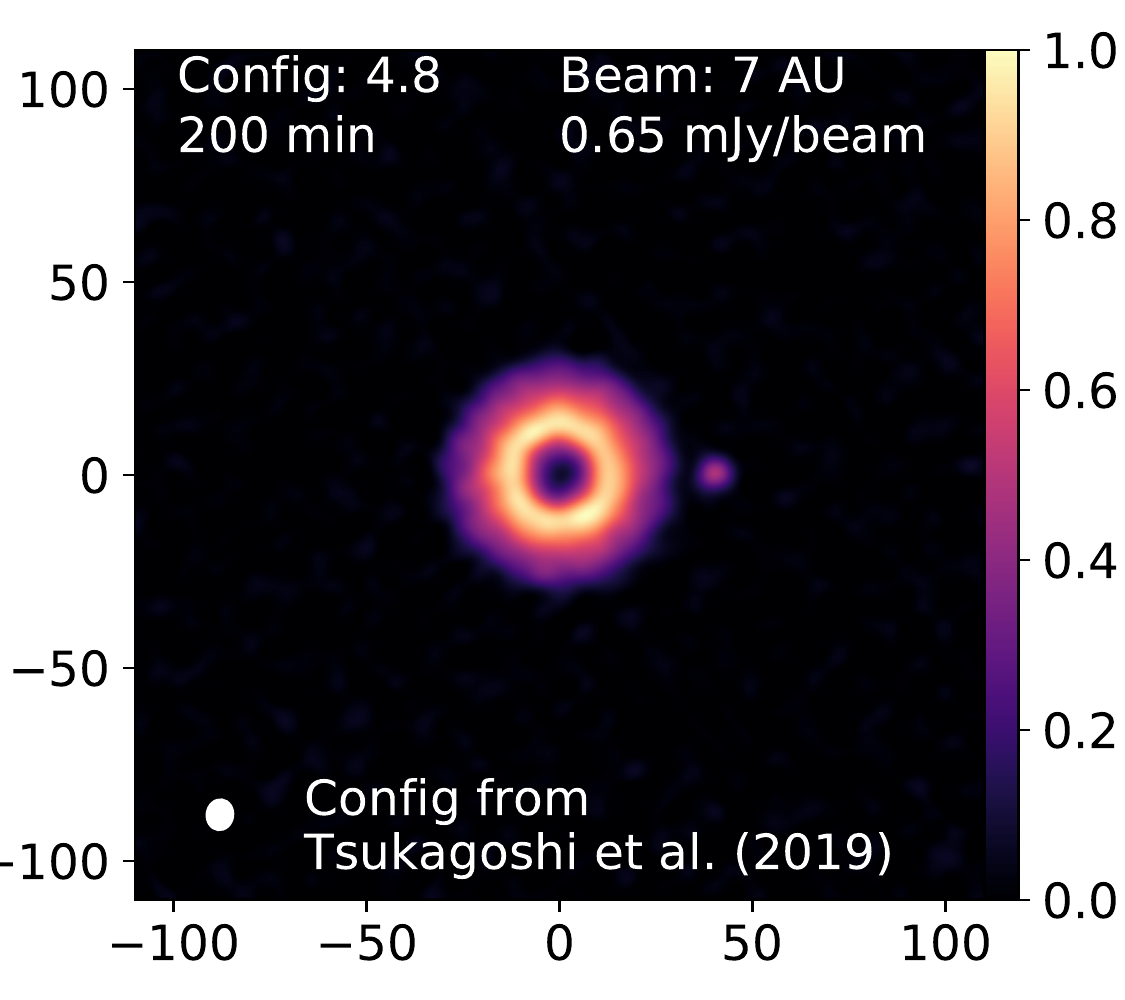}\\
    a) & b) & c) \\    
\end{tabular}
\caption{a) Surface density of dust in the system for a protoplanet with a radius of 1.5 AU around a 1 $\mj$ disc with a gas to dust ratio of 100. b) Low resolution \textsc{CASA} output for the parameters used in the ODISEA survey \citep{CiezaEtal19}. c) High resolution \textsc{CASA} output for the ALMA configuration from \protect\cite{TsukagoshiEtal19}. Only wide antenna configurations can resolve the protoplanet, the corresponding beam size is shown in the bottom left of each image. Since the protoplanet is optically thick it is dim compared to the disc despite having a surface density profile many orders of magnitude greater.}
\label{fig:surveys}
\end{figure*}

Although very little in our paper depends on the presence of a protoplanetary disc, its inclusion in the simulated images below is helpful to judge the ALMA detectability of GI planets as opposed to the discs. Recent ALMA surveys \citep{WilliamsEtal19} indicate that typical mm-grain dust mass of Class I and II protoplanetary discs is on the order of an Earth Mass.
In this work we consider a low mass disc with $M_D = 1 M_J$, an inner radius of 10 AU, an outer radius of 30 AU and a surface density profile of $\Sigma(r) = \Sigma_0$ (10 AU / $r$),  corresponding to a $\Sigma_0$ of 7.1 g/cm$^{-2}$. Taking a gas to dust ratio of 100, this leaves a dust mass of 3.18 $M_{\oplus}$.
The disc has a temperature profile $T(r) = 20K (100 AU/r)^{1/2}$, appropriate for a young star with a radius of $R_{\odot}$, a surface temperature of 4000 K and a corresponding luminosity of $\sim L_{\odot}/4$. 

The parameter space for the dust distribution in a realistic protoplanetary discs is enormous, covering both spatial and grain size dimensions, as well as grain composition. Covering this parameter space is impractical and we limit ourselves here to a simple toy model of a dust disc. This does not influences our results significantly since we are focused on the planet detectability, and in fact favour face-on discs.
Figure \ref{fig:setup} shows a graphical representation of our setup.
We consider two dust populations: vertically settled large grains and small dust that we assume follows the spatial distribution of the gas.  
We suppress the scale height of large dust in the disc by a factor of ten to represent dust settling, though we do not do this for dust inside the protoplanet in order to account for convection within its atmosphere \citep{HS08,HelledEtal08}.
We allocate the mass of these dust species assuming that the number distribution in grain size ($a$) is  $\textrm{dn/d}a \propto a^{-3.5}$ \footnote{This corresponds to 67\% of the dust mass in the top decade of grain size and 90\% of the dust mass being in the top two decades of the distribution.}, the standard power law exponent of the ISM \citep{MathisEtal77}. We set a minimum dust size ($a_{min}$) to 0.05 $\mu m$, a boundary dust size between small and large dust ($a_{mid}$) to 0.1 mm and then take a maximum dust size ($a_{max}$) of 1 mm. 

In interpreting millimeter dust gaps due to planets in ALMA data, \cite{LodatoEtal19} took a gap width of 5.5 times the planet's Hill Radius ($R_H$), based on hydrodynamical simulations by \cite{ClarkeEtal18}. This gives a gap width of 

\begin{equation}
    \Delta Gap = 5.5 a \left( \dfrac{q}{3} \right)^{1/3}
\end{equation}{}

where $a$ is the orbital separation of the protoplanet and $q$ is the mass ratio of the protoplanet to the central star. It follows that in order to resolve a protoplanet at the outer edge of a disc the required beam width of the observation should be less than the gap width to account for truncation of the disc by the protoplanet. For a protoplanet at 40 AU with a $q$ of $10^{-3}$, this corresponds to a length-scale of $\sim$ 15 AU. We represent this graphically in Figure \ref{fig:PP_lengthscale}, systems above the white lines are resolvable from the outer disc edge at each specified antenna configuration\footnote{This figure also applies to resolving circumplanetary accretion discs around post-collapse planets.}. 

While this calculation represents a minimum beam size, protoplanets surviving the initial GI phase may be flung to much wider orbits. However, viscous spreading of the disc may cause it to expand until it reaches and is truncated by the surviving protoplanet. 
We therefore suggest that protoplanet detection surveys should be optimised to detect protoplanets at the disc edge, requiring a minimum resolution of 15 AU. A resolution of less than 1 AU would be needed in order to resolve protoplanets directly.

\subsection{Radiative transfer}

\begin{figure*}
\begin{tabular}{ccc}
    \includegraphics[width=0.66\columnwidth]{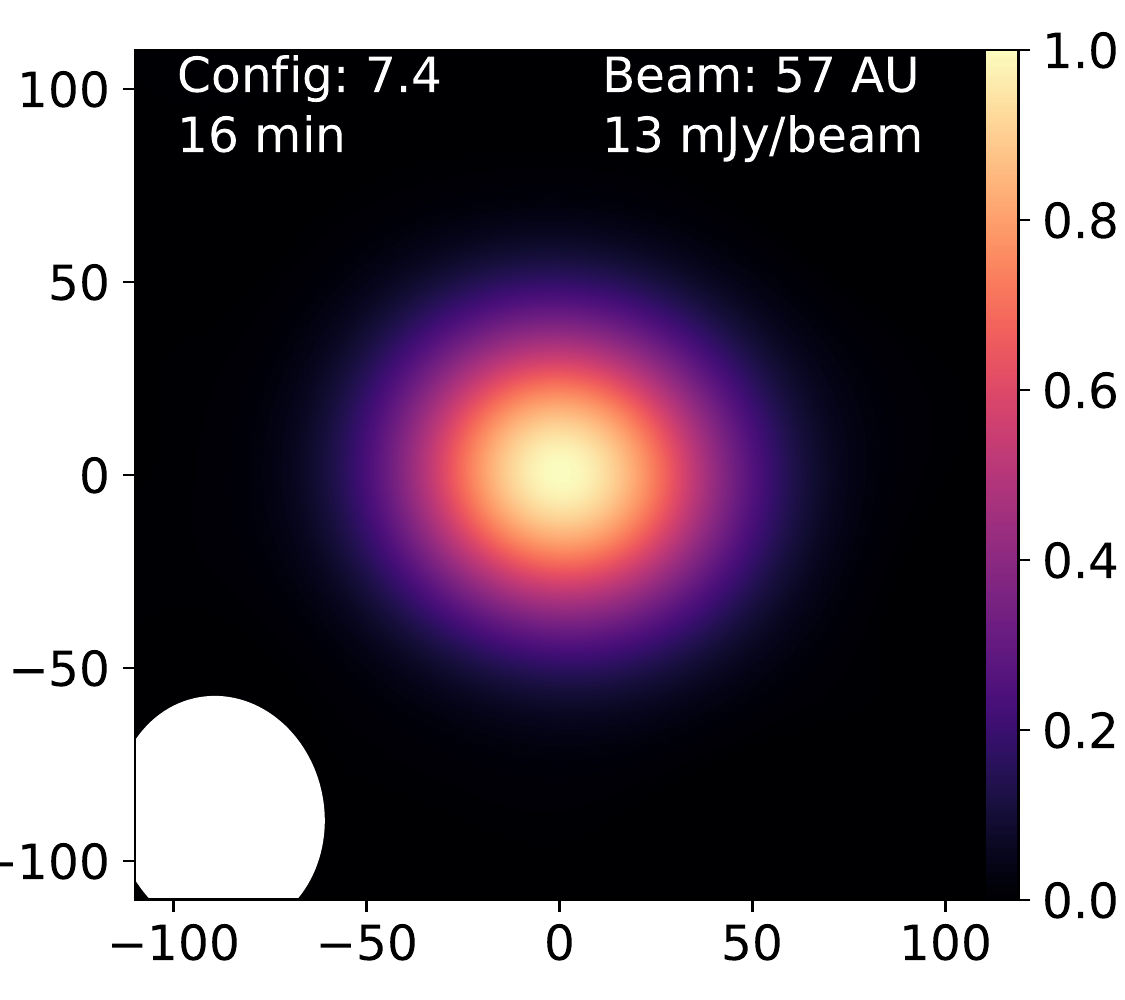} &
    \includegraphics[width=0.66\columnwidth]{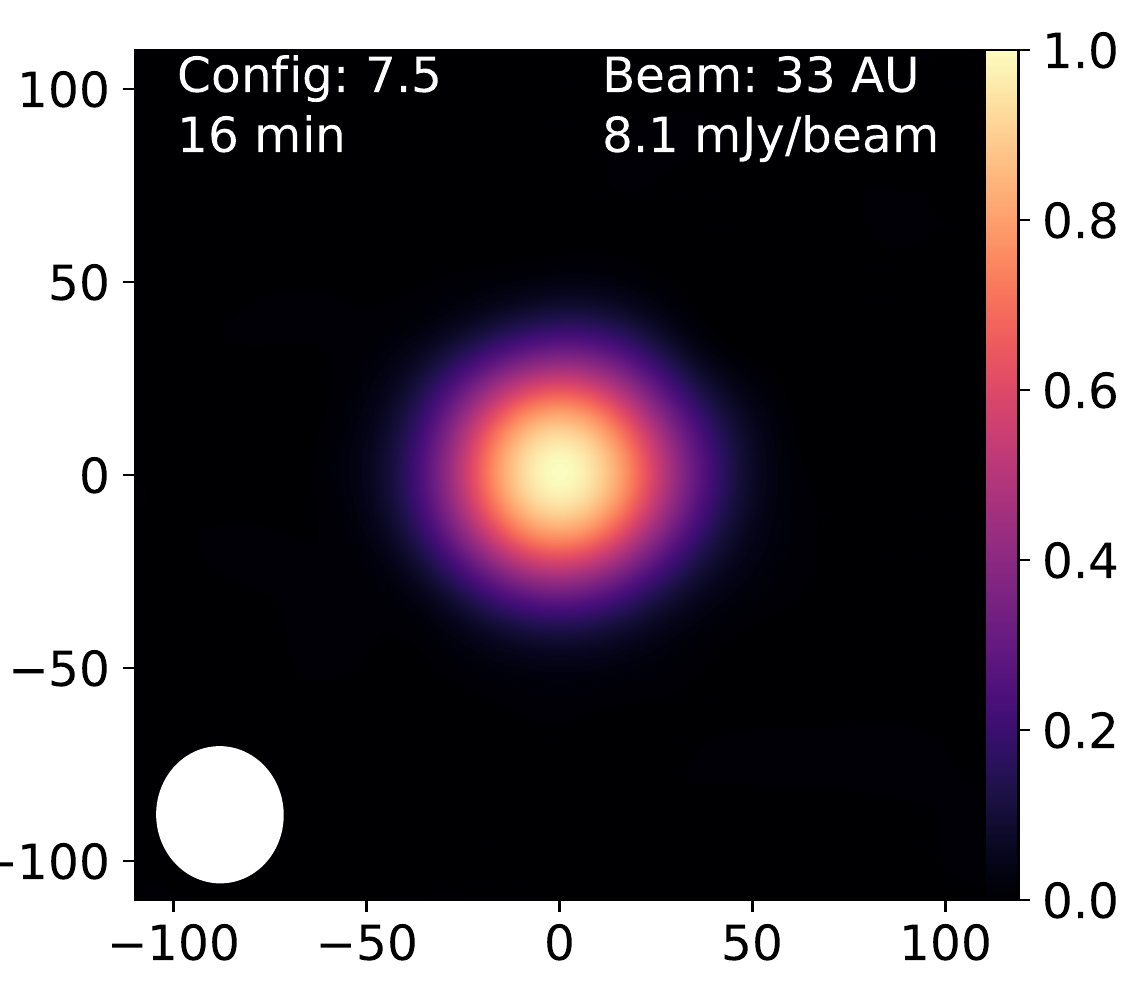} &
    \includegraphics[width=0.66\columnwidth]{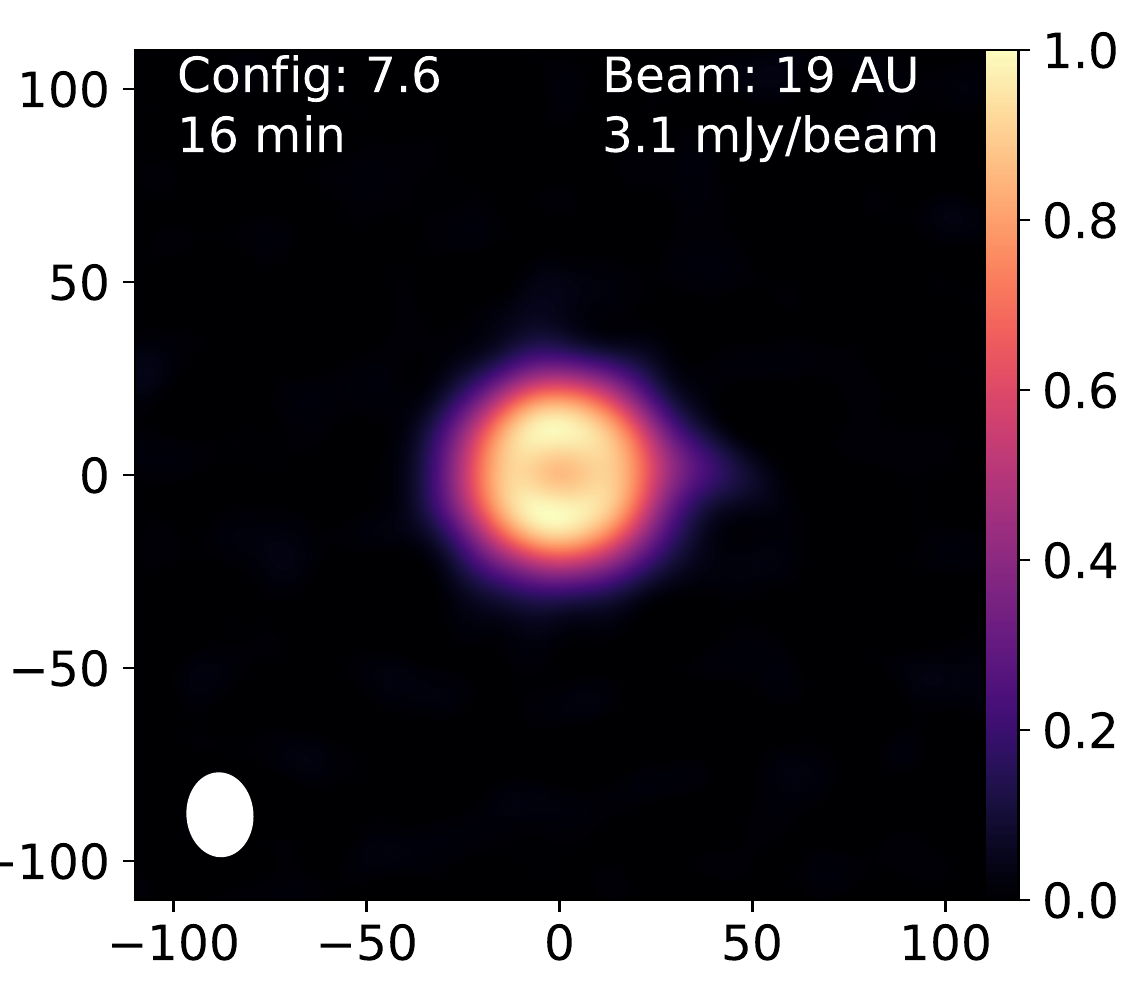}\\
    a) & b) & c) \\    
    \includegraphics[width=0.66\columnwidth]{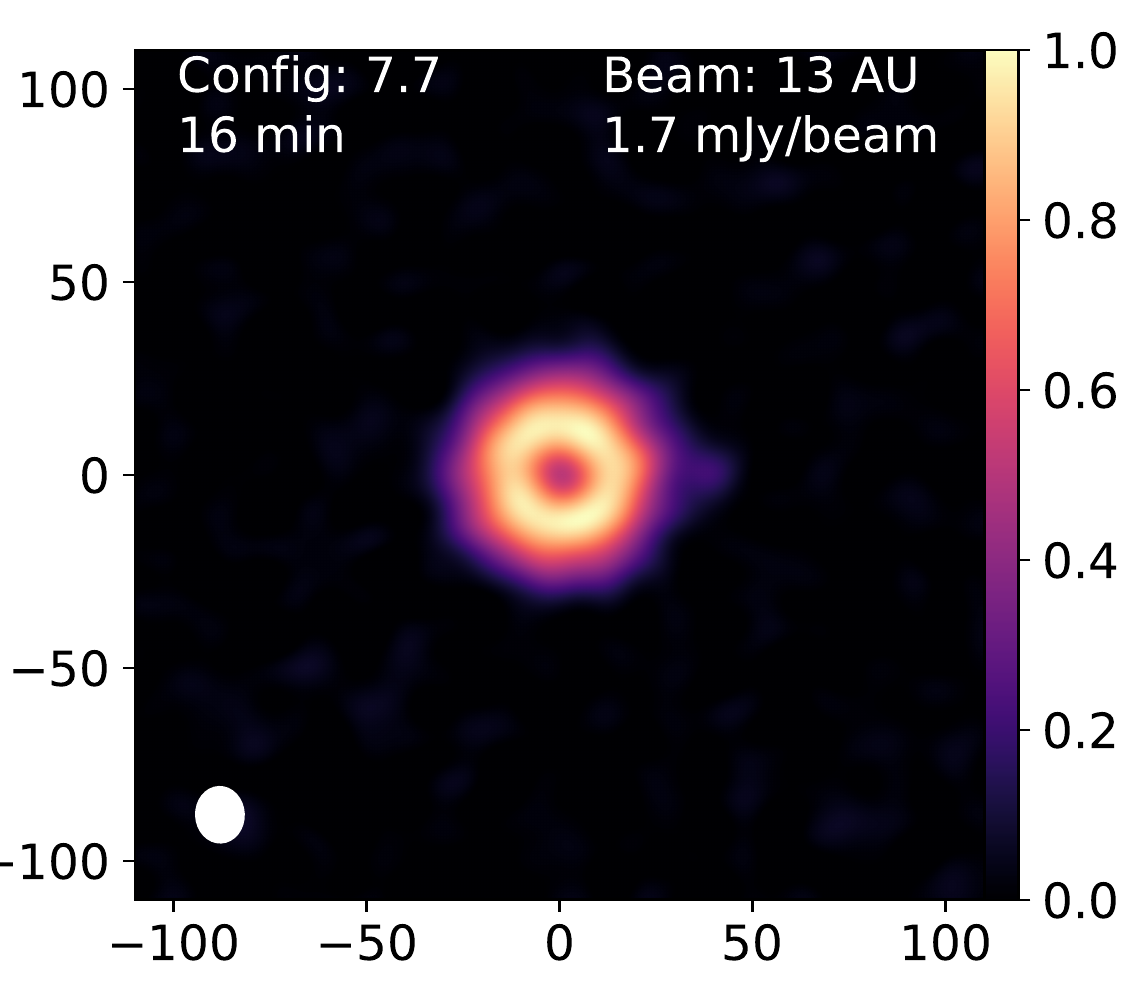} &
    \includegraphics[width=0.66\columnwidth]{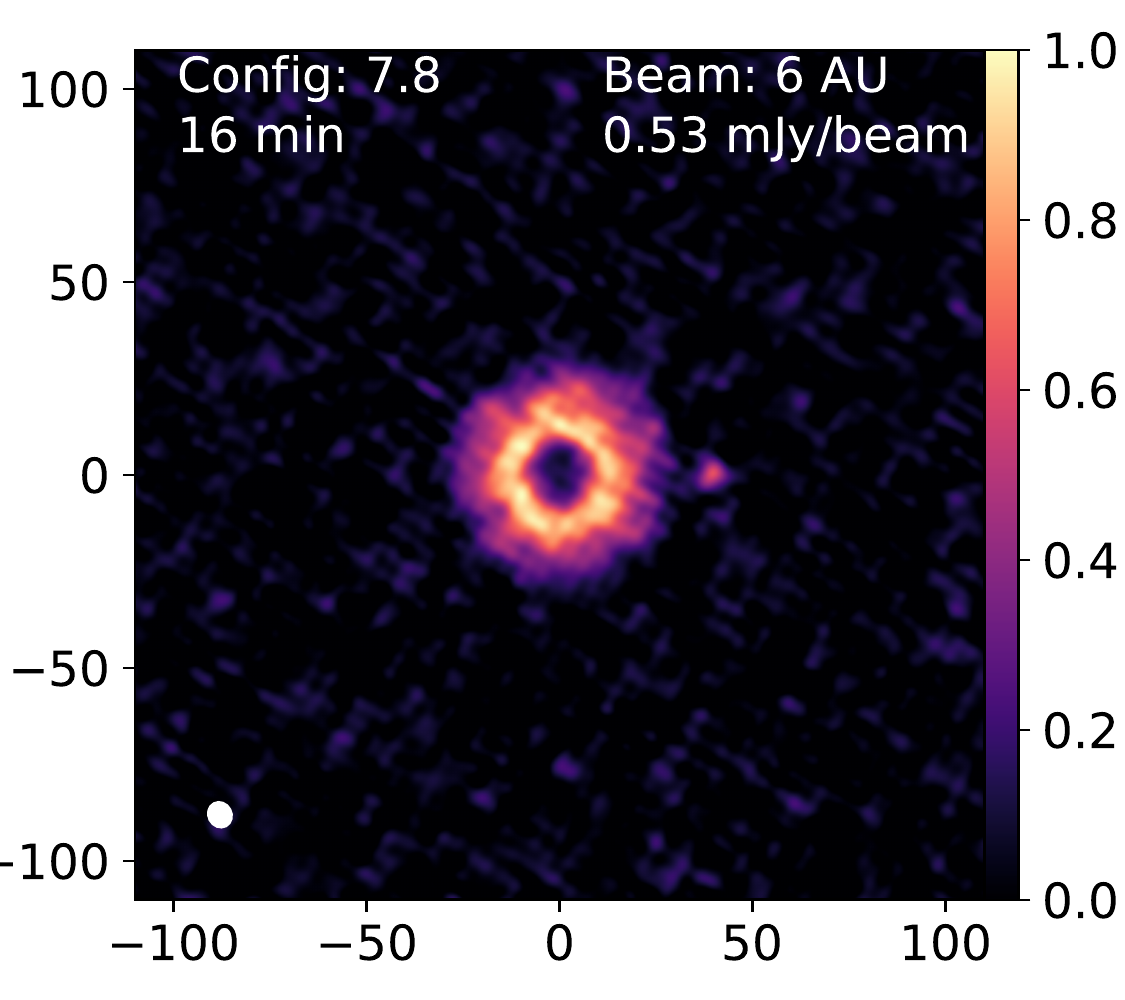} &
    \includegraphics[width=0.66\columnwidth]{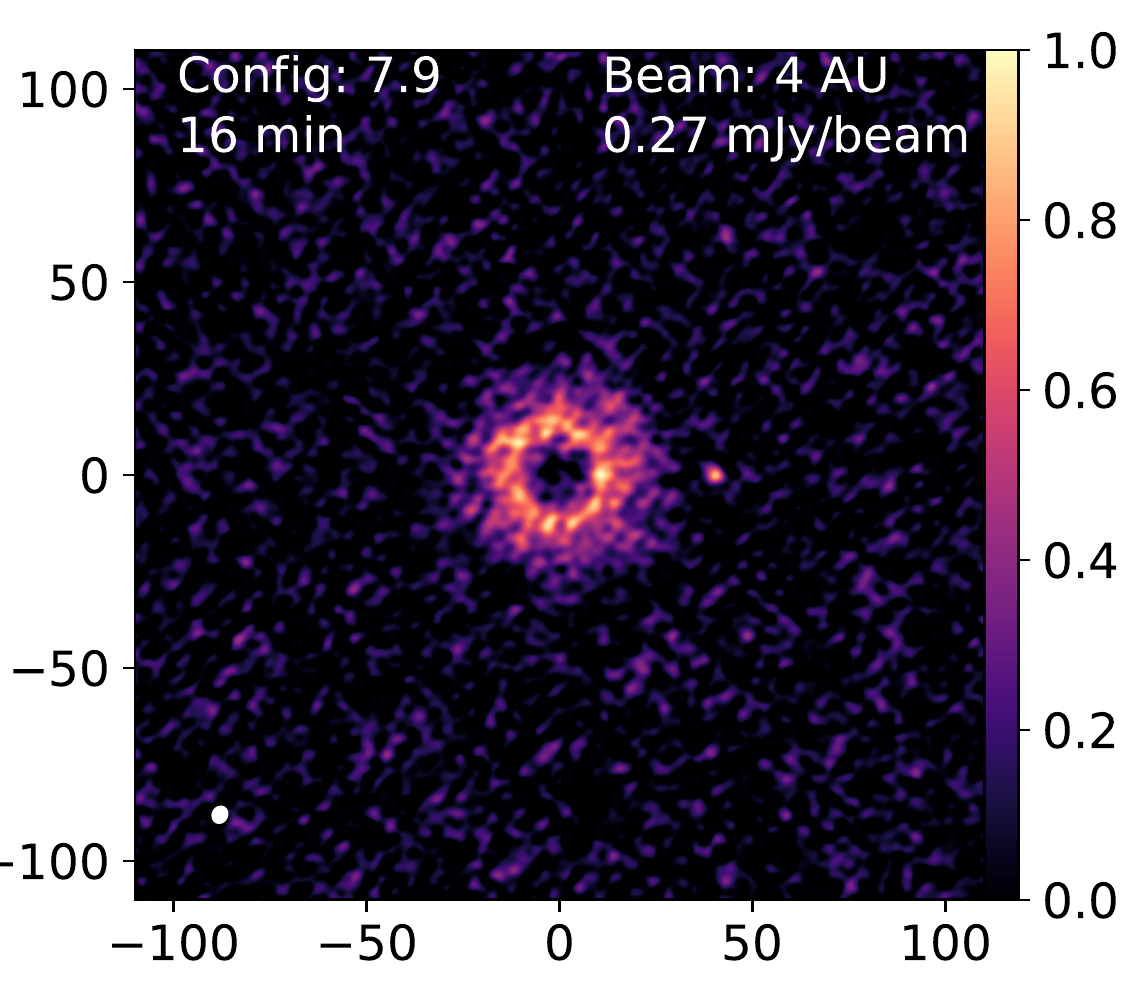}\\
    d) & e) & f) \\
\end{tabular}
\caption{Simulated ALMA images for a protoplanet with a radius of 1.5 AU at (40,0) AU around a 30 AU, 1 $\mj$ disc, shown for a variety of cycle 7 antenna configurations. The image scale is normalised to the maximum value of spectral flux density [mJy/beam] in each image and the signal-to-noise (SNR) ratio is defined as the flux at the location of the protoplanet divided by the standard deviation of the background noise.}
\label{fig:ALMA_cycles}
\end{figure*}

\begin{figure*}
\begin{tabular}{ccc}
    \includegraphics[width=0.66\columnwidth]{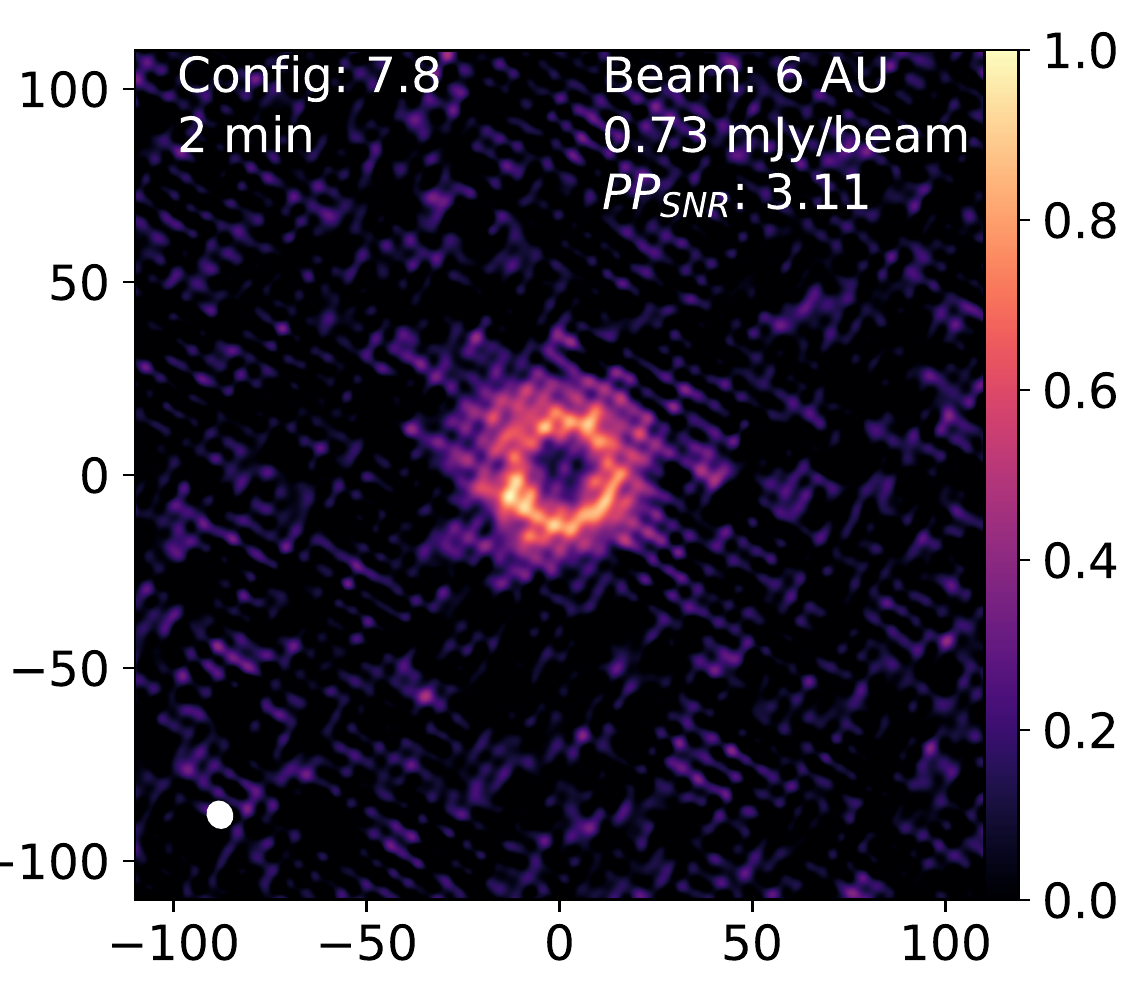} &
    \includegraphics[width=0.66\columnwidth]{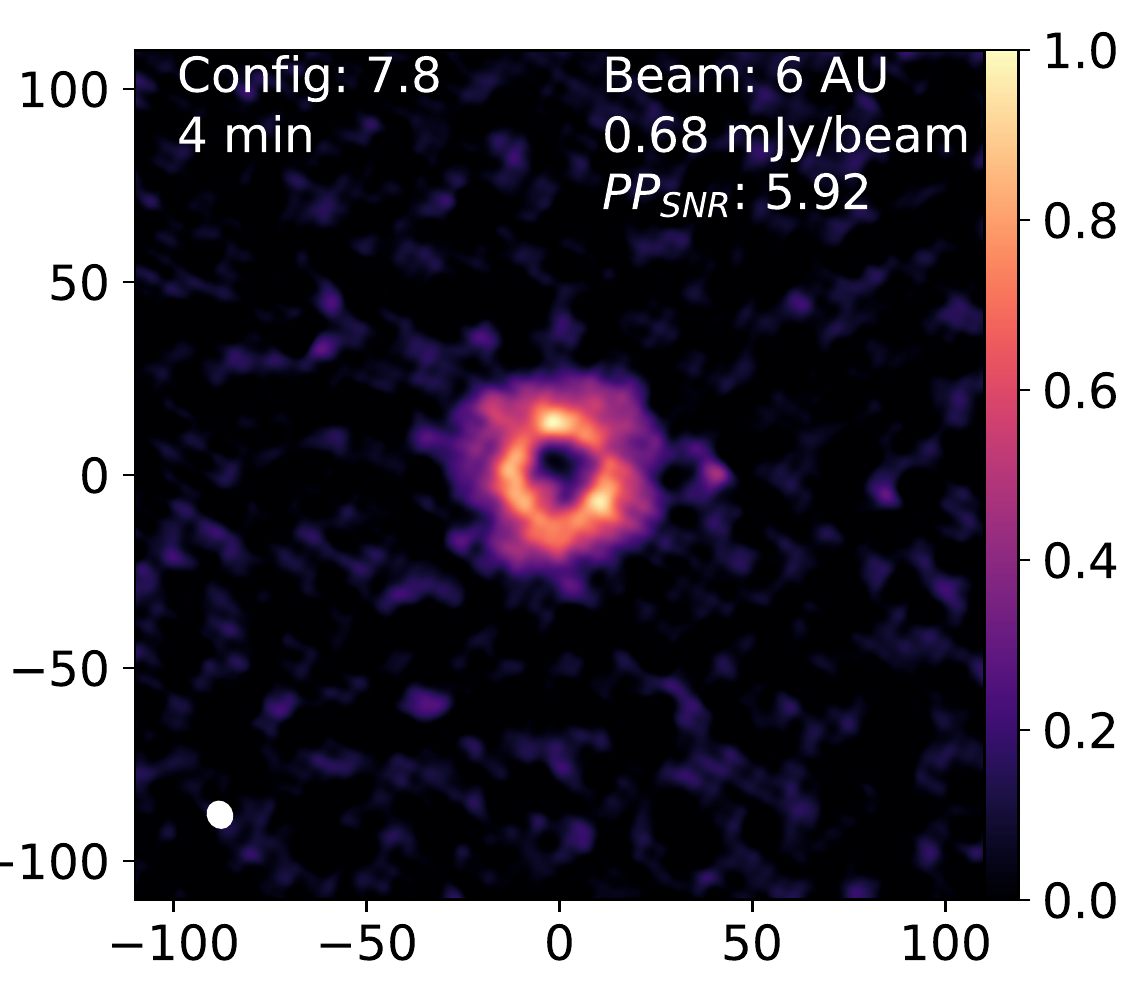} &
    \includegraphics[width=0.66\columnwidth]{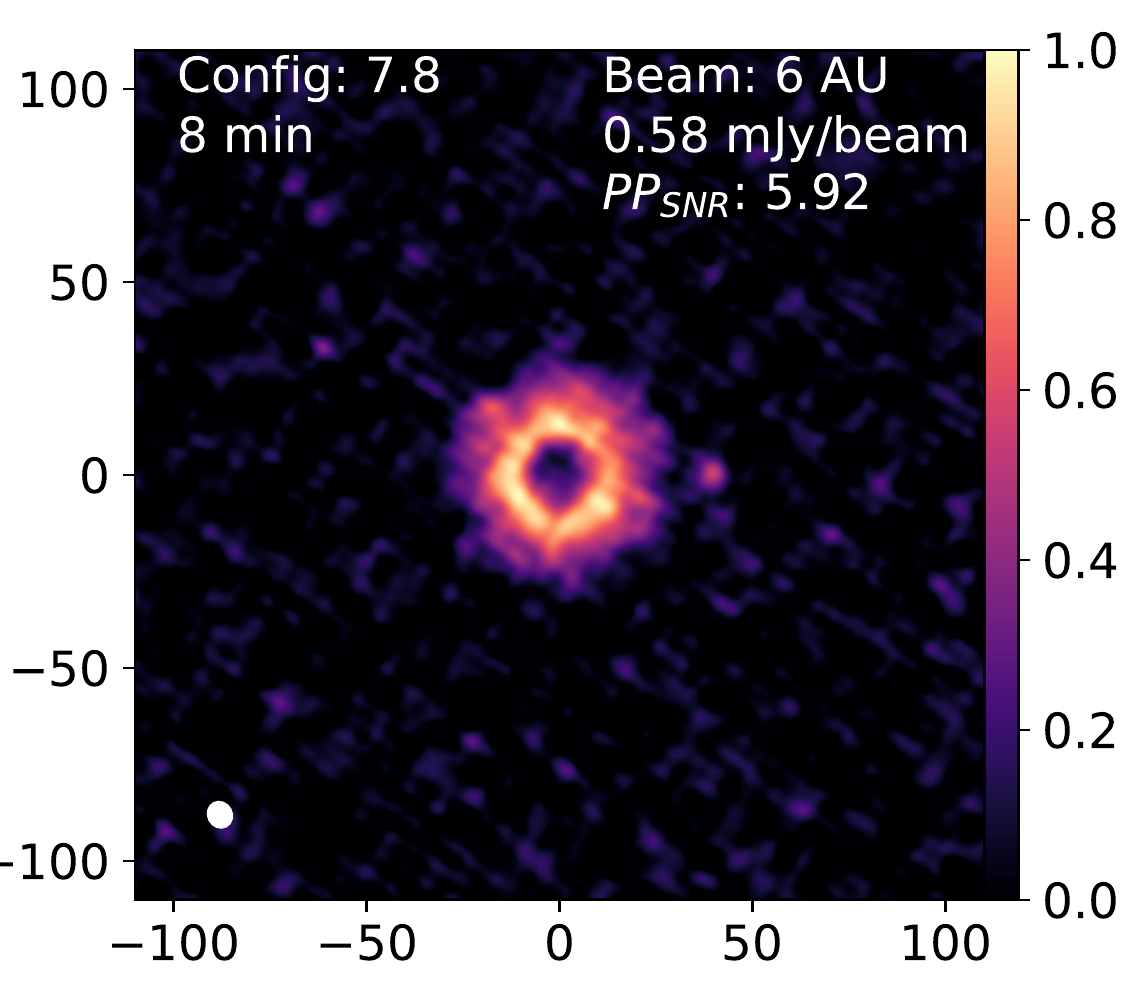}\\
    a) & b) & c) \\    
    \includegraphics[width=0.66\columnwidth]{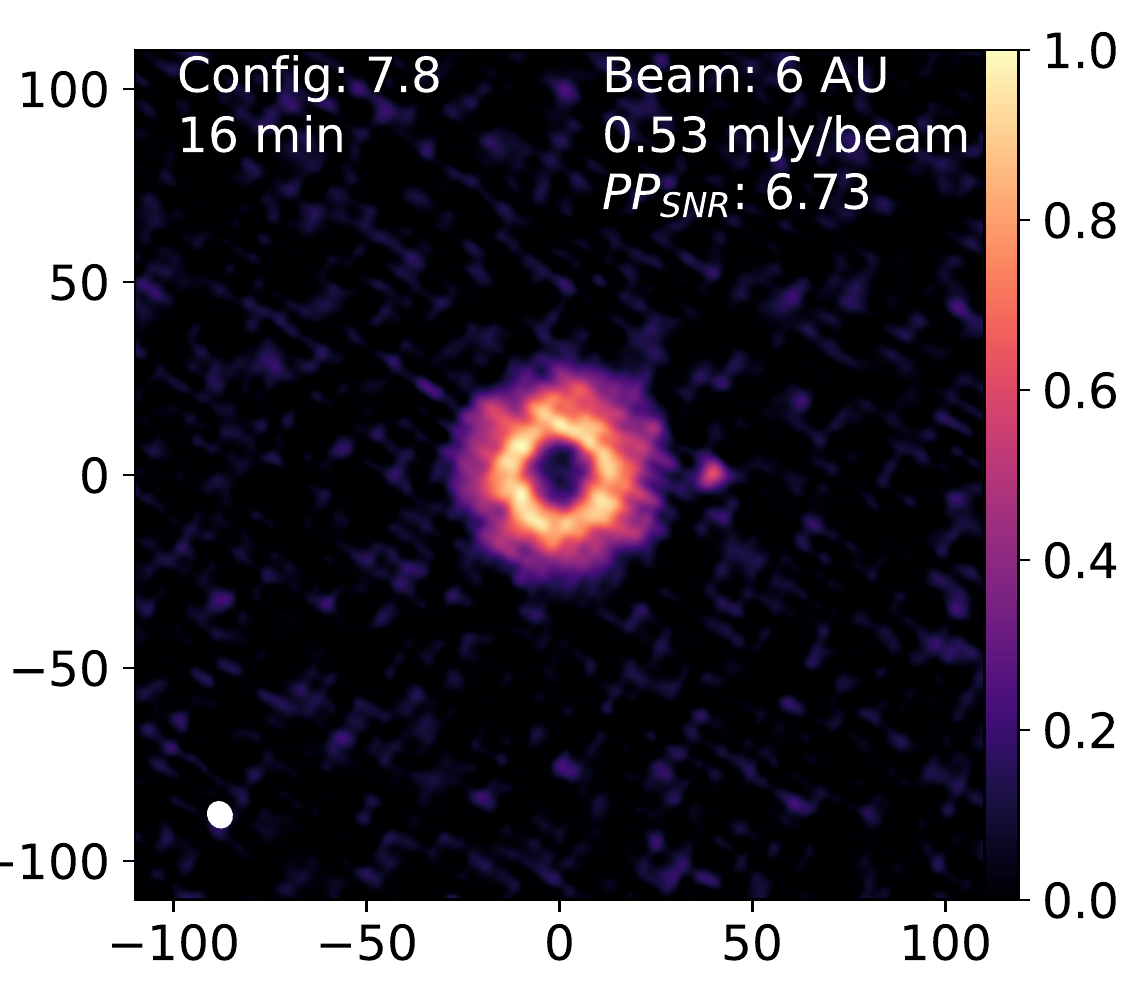} &
    \includegraphics[width=0.66\columnwidth]{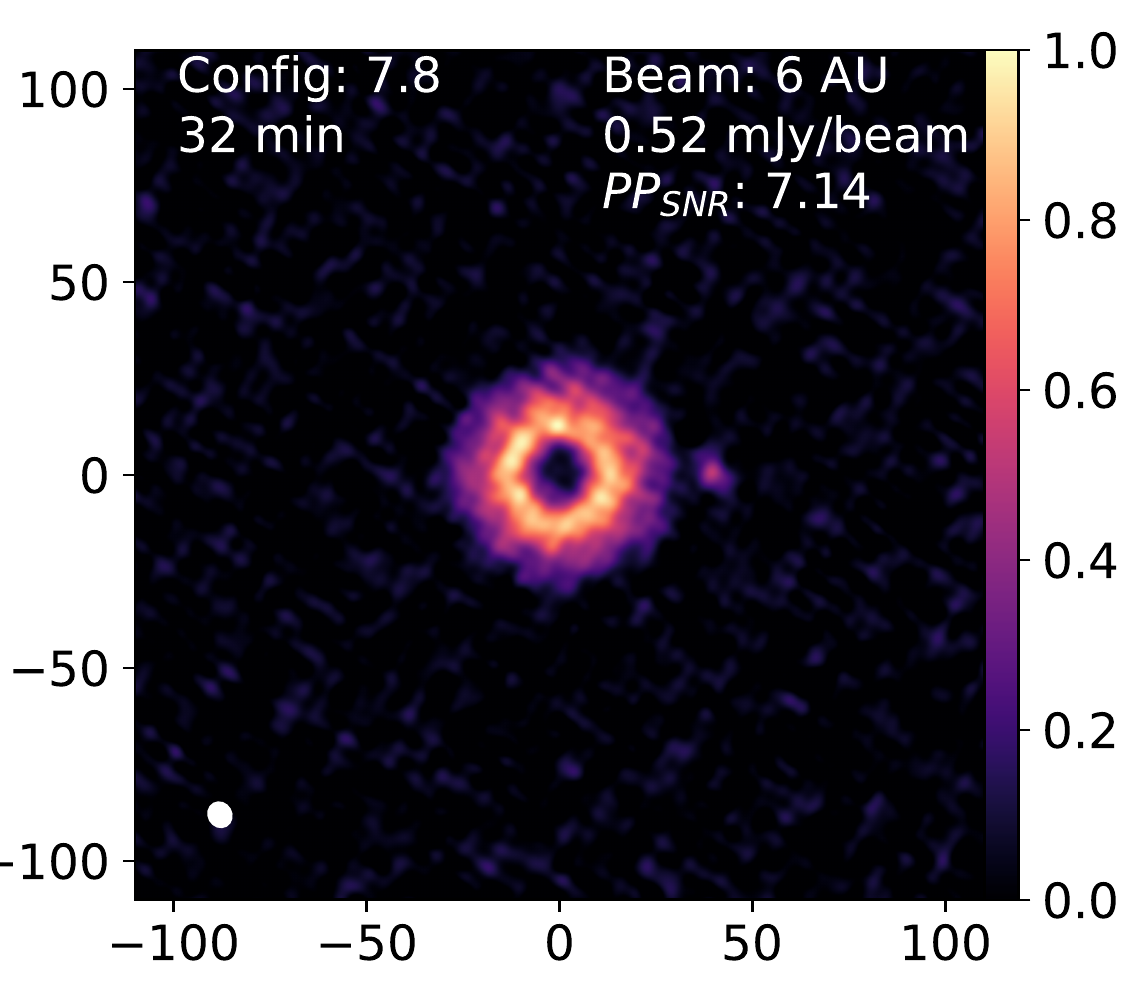} &
    \includegraphics[width=0.66\columnwidth]{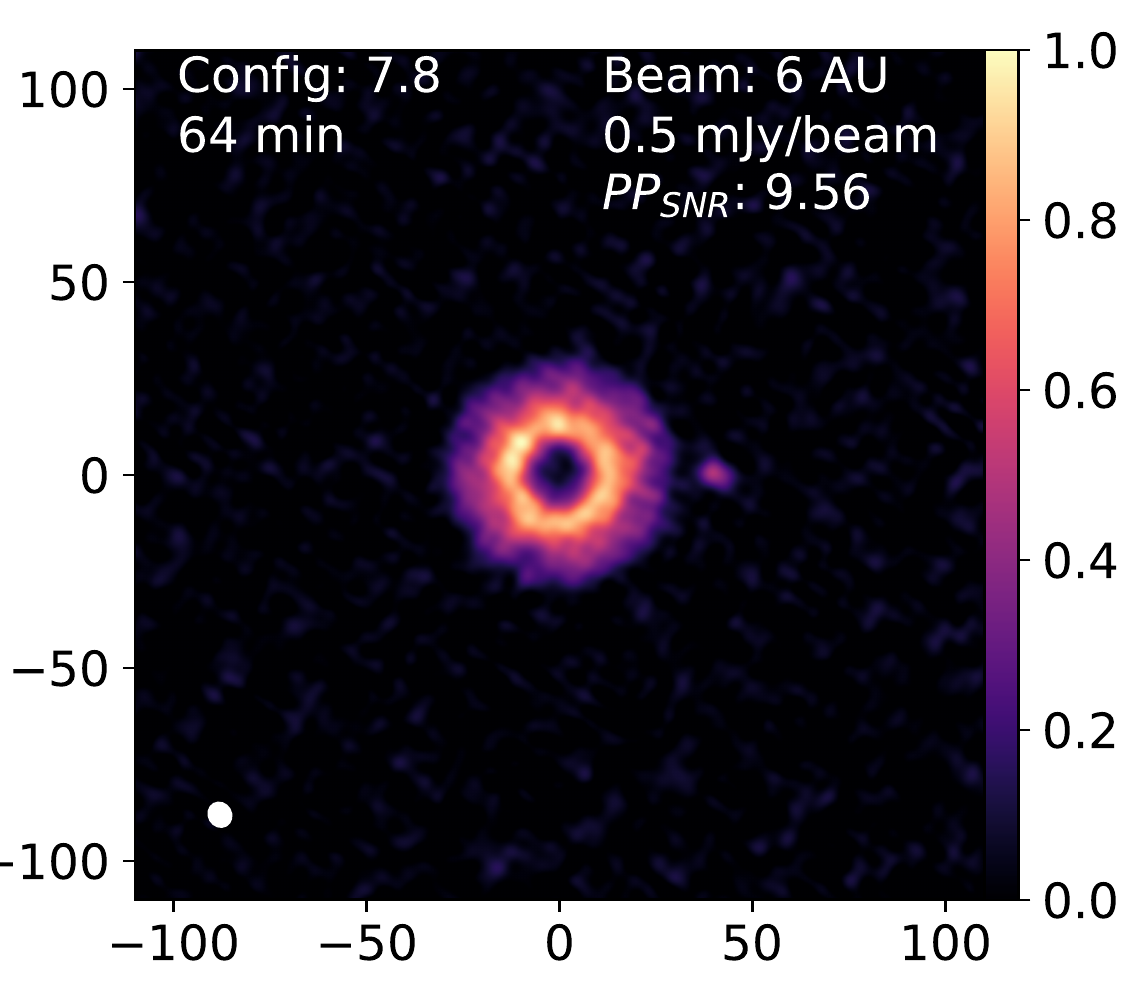}\\
    d) & e) & f) \\
\end{tabular}
\caption{Simulated ALMA images for a protoplanet with a radius of 1.5 AU at (40,0) AU around a 1 $\mj$ disc, same as Figure \ref{fig:ALMA_cycles} but now shown for the cycle 7.8 configuration over a variety of integration times. Increasing the integration time increases the signal to noise of the detection, recorded in the top right of each image. Protoplanets may be as compact as 0.1 AU depending on evolution model, in which case the SNR values would be considerably lower.}
\label{fig:ALMA_ints}
\end{figure*}

We use the radiative transfer code \textsc{RADMC-3D} \citep{DullemondEtal12} to produce an image of the continuum emission from our system using our preset gas temperatures and assuming that gas and dust are thermally coupled. We consider emission from a star with a surface temperature of 4000 K, consistent with the irradiated temperature profile that we set for the disc. 
We made the choice to use preset gas temperatures since only the surface temperature of a protoplanet is affected by external irradiation, deeper layers and even the contraction rate are only weakly affected by external heating \citep{VazanHelled12}\footnote{At higher levels of irradiation, when the irradiation temperature approaches $\sim 1/3$ of the central temperature, the protoplanet becomes unbound and is disrupted, see Section 4.1.2 in \cite{Nayakshin15a} and \cite{VazanHelled12}.}. 

We produce an adaptive mesh grid with a resolution that follows the density profile of the gas. This leads to a cell size of 0.1 AU at the location of the protoplanet.
We set our dust opacity based on a mixture of 60\% silicates, 15\% amorphous carbon and 25\% porous empty space. These are the default parameters of  \textsc{Opacity tool}, a software tool designed to compute opacity tables that was an outcome of the DIANA project for modelling dust in protoplanetary discs \citep{WoitkeEtal16}.
Our disc is optically thick at 1.3 mm (ALMA Band 6) from it's inner edge at 10 AU out to 20 AU.

Due to the simple geometric nature of our set up and the optical thickness of our protoplanet, we found that our calculated SNR values converged using only $10^5$ photon packets in \textsc{RADMC-3D} (Appendix \ref{app:conv}). We opted to use $10^7$ photon packets in this paper in order to produce smooth images for presentation. The discs are optically thick in the midplane out to 20 au.

\subsection{Simulating ALMA observations}

To account for interferometric effects in our synthetic observations we use the ALMA simulator from the \textsc{CASA} software (ver 5.5.0) \cite{McMullinEtal07} to process the \textsc{RADMC-3D} emission maps. Following \cite{HallEtal19}, we corrupt the visibilities with thermal noise using the Atmospheric Transmission at Microwaves (\textsc{ATM}) code \citep{PardoEtal01}.

Recent ALMA surveys have constrained numerous properties of the discs in local clusters \citep{AnsdellEtal17}, although they have not yet reached sufficient resolution to detect wide-orbit GI planets. 
To demonstrate this, we first reproduce the setup of the Ophiuchus DIsc Survey Employing ALMA \cite[ODISEA,][]{CiezaEtal19} to quantify the difficulty of detecting protoplanets with low resolution observations.
This survey sampled the 147 brightest discs in the Ophiuchius Molecular Cloud \citep[d = 140 pc, ][]{OrtizLeonEtal17} at a resolution of 28 AU/0.2" with a typical RMS of 0.15 mJy. It was conducted in ALMA Cycle 4 on July 13/14th 2017 in Band 6 (1.3 mm/230 GHz), with an antenna configuration of C40-5. To approximate their continuum observations we take a spectral window centred at 225 GHz with a bandwidth of 3.98 GHz. We set the total integration time to 60 seconds\footnote{\cite{CiezaEtal19}, private communication.} and set the precipitable water vapor (PWV) value to 1.5.

The result of this setup can be seen in the middle panel of Figure \ref{fig:surveys}. Unsurprisingly, it is impossible to detect the protoplanet in this configuration since the ODISEA survey was designed to quickly sample many discs at a low resolution. The  protoplanet vanishes when convolved with the disc due to the large beam size. This occurs despite the protoplanet being well separated from the outer edge of the disc. The planet radiation {\em intensity} is also intrinsically much higher than the intensity at the outer disc edge, but since the planet emitting area is small compared to that of the disc, its overall luminosity is insufficient for the planet to be discerned by ALMA in this configuration.

ALMA is however capable of operating at much higher resolution than the ODISEA survey. For instance, \cite{TsukagoshiEtal19} conducted a recent Band 6 observation of TW Hya using the configuration C43-8 with an integration time of 200 minutes, which contains tentative evidence for a protoplanet in the disc at 50 au from the central star \citep{nayakshinetal2020}. The far right panel of Figure \ref{fig:surveys} shows how adopting these parameters improves our image, the protoplanet is now clearly resolvable.
In the following section we attempt to quantify the ideal antenna configuration and integration times required in order to resolve long-lived protoplanets around compact protoplanetary discs, should they exist. We use a shorthand to keep notation compact: `Cycle 7.7' corresponds to a synthetic observation made in ALMA's cycle 7 in the C7 configuration.

\section{Results}
\label{sec:results}

\subsection{Required resolution and sensitivity}

In order to quantify the observability of our protoplanets we calculate their signal to noise ratio (SNR), defined as the ratio of the flux at the location of the protoplanet to the standard deviation of the background flux. 

Figure \ref{fig:ALMA_cycles} shows how the choice of ALMA antenna configuration affects synthetic observations of our $R_{PP}$ = 1.5 AU protoplanet at 40 AU for a fixed integration time of 16 minutes. As demonstrated in Figure \ref{fig:PP_lengthscale}, the cycle 7.4, 7.5 and 7.6 configurations are unable to resolve the protoplanet from the outer edge of the disc. Meanwhile, the protoplanet is resolvable for the larger baseline configurations even after 16 minutes. 
While larger beam sizes can resolve protoplanets at wider separations, it is certainly worth considering protoplanets close to the disc edge, since the outer parts of a disc are the most likely to become gravitationally unstable. 
Futhermore, protoplanets are massive enough to truncate the disc and so even wide-orbit protoplanets may find themselves at the outer disc edge after a period of viscous expansion. Nevertheless, in the following sections we expand this analysis to consider protoplanets at wider orbits.

In Figure \ref{fig:ALMA_ints} we explore the effect of varying the total integration time for the cycle 7.8 configuration from Figure \ref{fig:ALMA_cycles}. As expected, at low integration times the image is dominated by background noise. Visually, it is possible to distinguish the protoplanet from the disc after only 4 minutes, though the image becomes clearer with longer integrations. 

Noise remains a major contributing factor to the image, demonstrated by the blotchy disc even in the 64 minutes observation. In terms of a reasonable comparison study, \cite{TsukagoshiEtal19} found an SNR value of 15 for their feature in TW Hya, though this is complicated slightly since that object is embedded within the disc. In this paper we treat systems with SNR values greater than five as detections, though such low contrast features are frequently passed over in typical disc surveys. 

It is common for surveys to concatenate different ALMA configurations in order to resolve both large and small features within discs. However, since the projected size of our protoplanets is universally smaller than the highest resolution ALMA configuration, we found that concatenation offered no improvement over using a single configuration (see Appendix \ref{app:concat}).

\subsection{SNR for young pre-collapse protoplanets}

\begin{figure*}
\includegraphics[width=1.0\columnwidth]{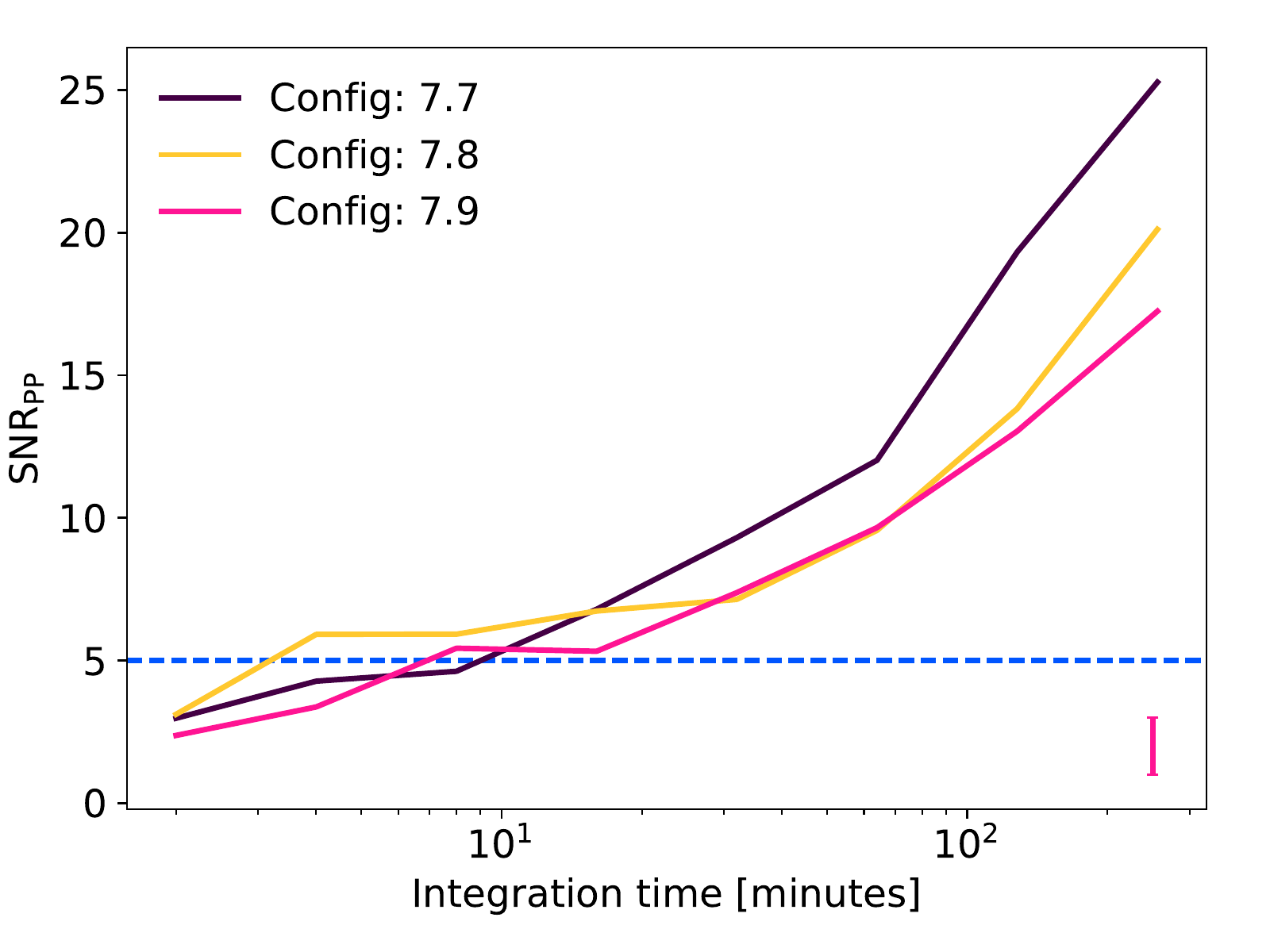}
\includegraphics[width=1.0\columnwidth]{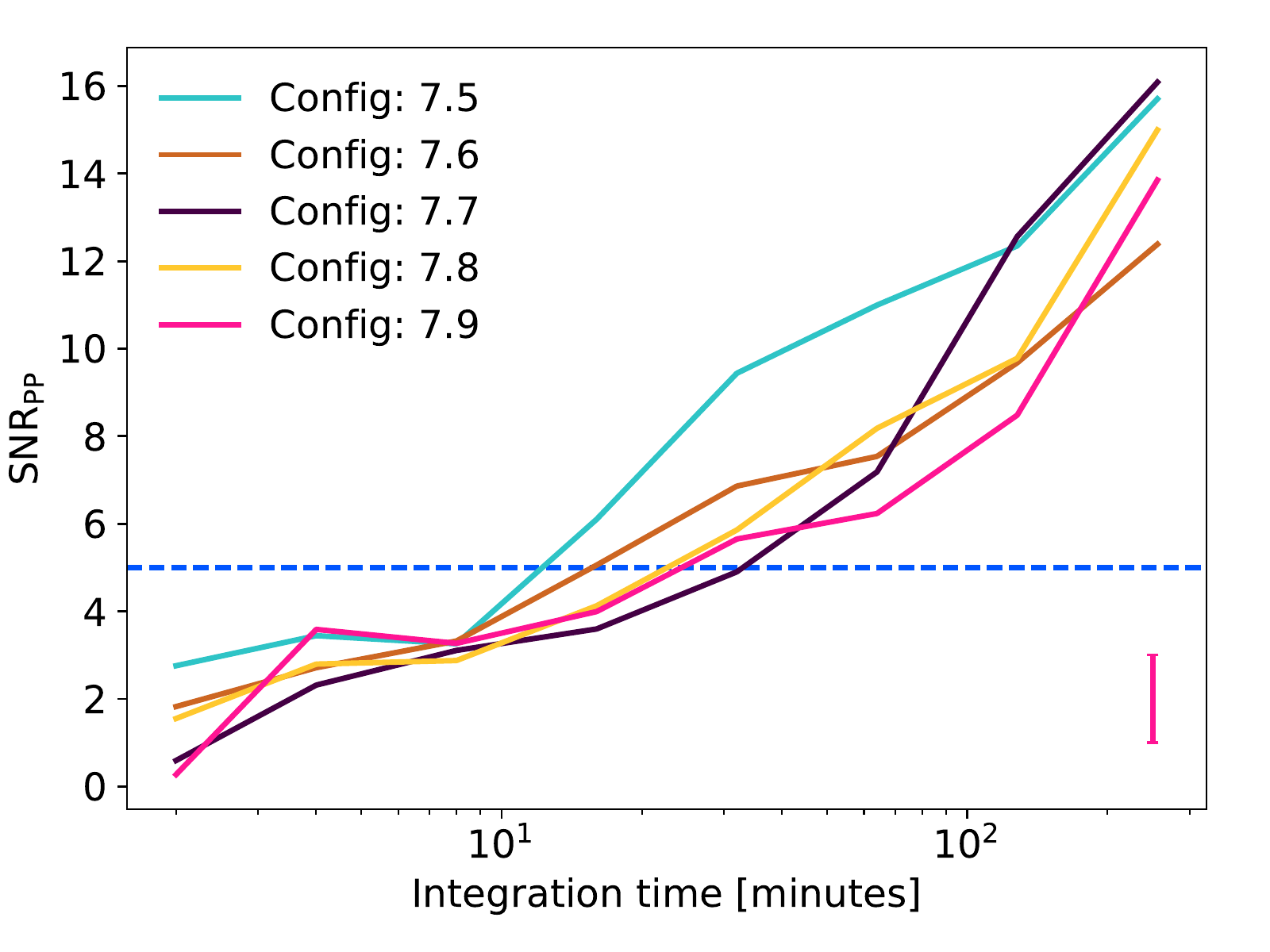}
\caption{Signal-to-noise ratios for a protoplanet with a radius of 1.5 AU at 40 AU (left) and 100 AU (right) for a variety of ALMA antenna configurations and total integration times of 2, 4, 8, 32, 64, 128 and 256 minutes.  The errorbar in the bottom right represents the +/- one standard deviation values of the SNR. Config 7.7 is shorthand for a synthetic observation made in ALMA cycle 7 in the C7 configuration.}
\label{fig:3MJ_SNR}
\end{figure*}

\begin{figure*}
\includegraphics[width=1.0\columnwidth]{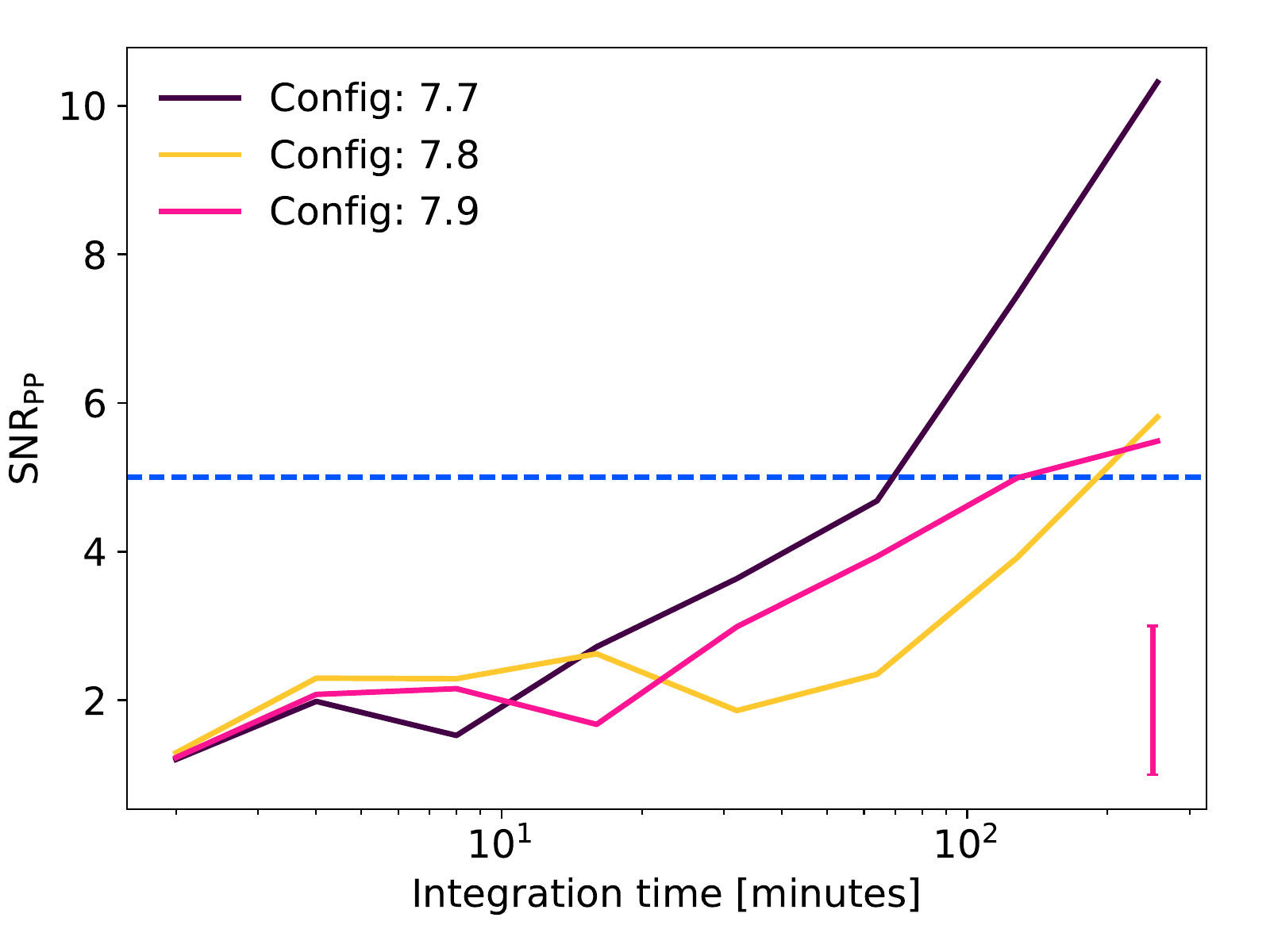}
\includegraphics[width=1.0\columnwidth]{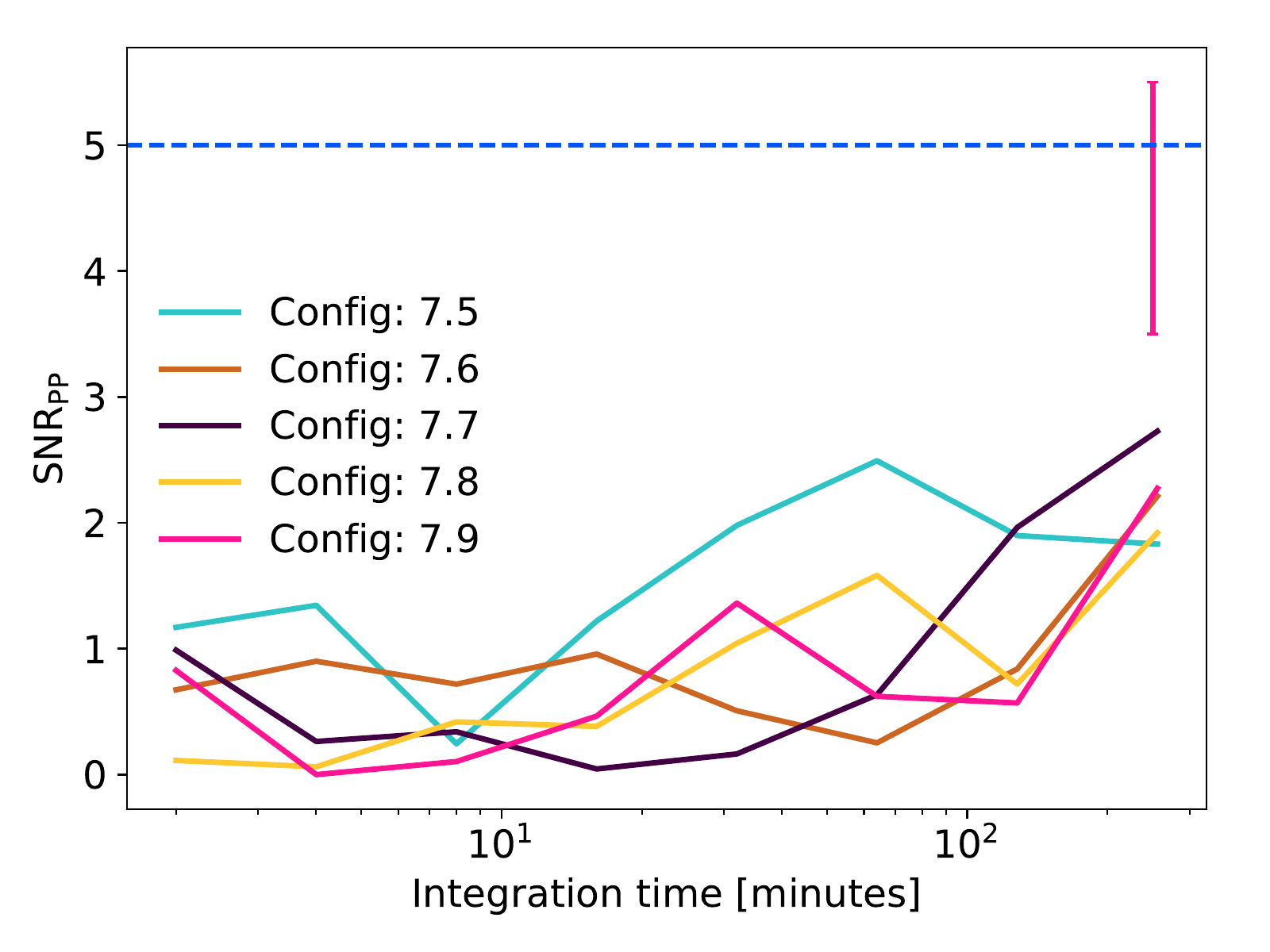}
\caption{Same as Figure \ref{fig:1MJ_SNR} but now for a protoplanet with a radius of 1 AU. More compact protoplanets are less luminous due to their smaller surface emission areas. Config 7.7 is shorthand for a synthetic observation made in ALMA cycle 7 in the C7 configuration.}
\label{fig:1MJ_SNR}
\end{figure*}

Figure \ref{fig:3MJ_SNR} plots SNRs for a protoplanet with a radius of 1.5 AU at a separation of 40 (left) and 100 (right) AU for a variety of ALMA antenna configurations and integration times. 
From the left panel we see that after 10 minutes of integration time the SNR of the protoplanet is consistently above 5, which we consider our threshold for detection.nIn the right hand panel we see that the SNR decreases for more a distant protoplanet. Due to our temperature prescription the protoplanet effective temperature is 20 K at 100 AU, though this may be optimistically high depending on stellar irradiation. In any case, the SNR of such a protoplanet only rises above 5 after $\sim$ 20 minutes of integration time. 

The SNR values broadly lie within one standard deviation of each other, indicated by the pink error bar in the bottom right of this figure. These tests show that antenna configuration has little impact on the SNR ratios, 
provided that the protoplanet is resolved from the outer disc edge. This result is expected since these protoplanets are essentially point-sources compared to the beam sizes considered here. Appendix \ref{app:int100} contains a version of Figure \ref{fig:ALMA_ints} for our 1.5 AU protoplanet at 100 AU, at this separation we use a cooler surface temperature of 20K in order to represent reduced stellar irradiation. This significantly reduces the SNR value of such a protoplanet.

\subsection{SNR for old pre-collapse protoplanets}

We now consider a protoplanet reaching the end of its pre-collapse phase that has cooled and contracted over time. We set a smaller radius of 1 AU, a reduction of 2.25 to the emitting surface area. An older protoplanet represents an ideal candidate for detection since it is challenging to observe very young, obscured systems. 

Neglecting any change in surface temperature, protoplanets with smaller surface emission areas are more challenging to detect. Figure \ref{fig:1MJ_SNR} plots the SNR for our 1 AU protoplanet, we see that it is consistently lower than the SNR of the younger protoplanet in Figure \ref{fig:3MJ_SNR}. The SNR of this  protoplanet only rises above 5 after $\sim$ 60 minutes of observing time. This detection problem only gets worse when the protopanets are colder. The right hand panel of Figure \ref{fig:1MJ_SNR} shows the case for a 1 AU protoplanet at 100 AU with a surface temperature of 20 K, the SNR value remains below 5 even after 300 minutes of observation.

We can see from the left panel of Figure \ref{fig:1MJ_SNR} that detecting protoplanets at the end of their pre-collapse phase with ALMA requires hundreds of minutes of observation time.
Even high resolution surveys with integration times of an hour such as the DSHARP programme \citep{DSHARP1} would miss a wide-orbit protoplanet with a radius of 1 AU and an effective temperature of $\sim$ 20 K. Such planets are hence much less detectable than the young, massive and close-orbit protoplanets, for which we advocated relatively short ($\sim$ 10 minutes) integration times in previous section.


\section{Protoplanet detection with ALMA}
\label{sec:obs}

\begin{figure*}
\includegraphics[width=1.0\columnwidth]{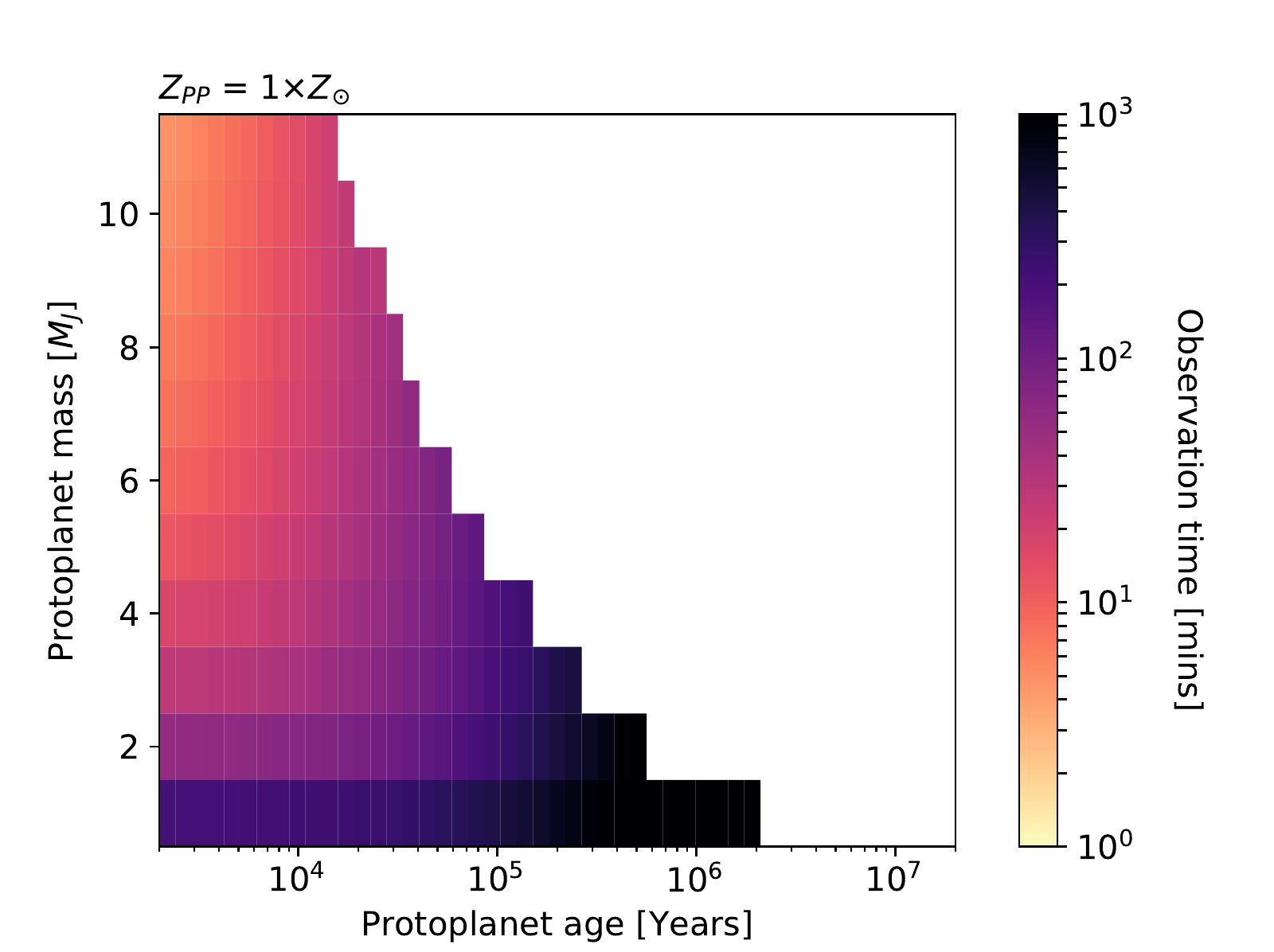}
\includegraphics[width=1.0\columnwidth]{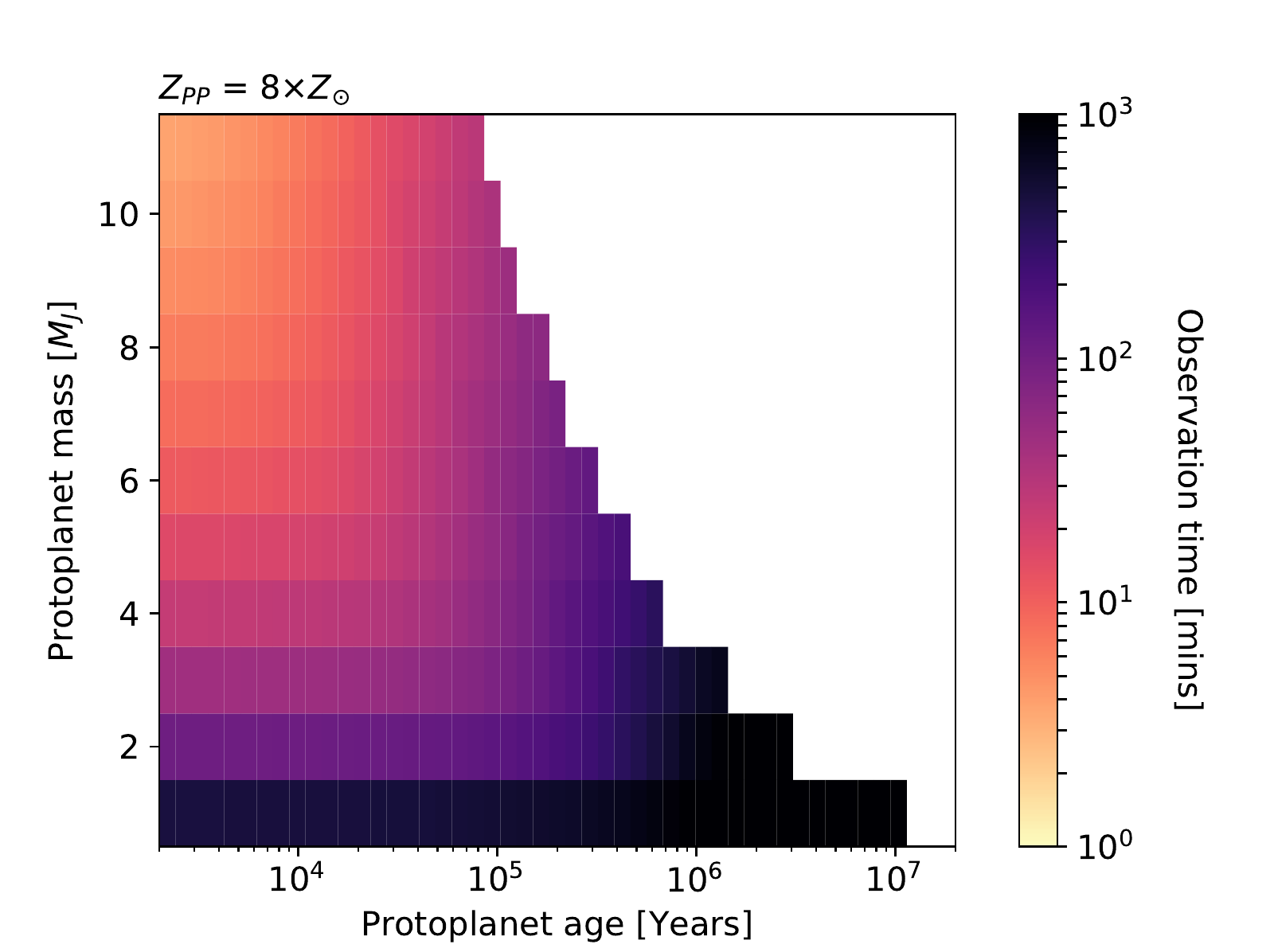}
\caption{The lifetime of different mass protoplanets coloured by required observation time. Left: $Z_{PP}=Z_{\odot} = 0.015$, right: $Z_{PP}=8 \times Z_{\odot}=0.12$. Metal enriched protoplanets spend longer in the pre-collapse configuration due to their higher opacities and correspondingly lower luminosities, but consequently require longer observation times.}
\label{fig:age_obs}
\end{figure*}

\begin{figure*}
\includegraphics[width=1.0\columnwidth]{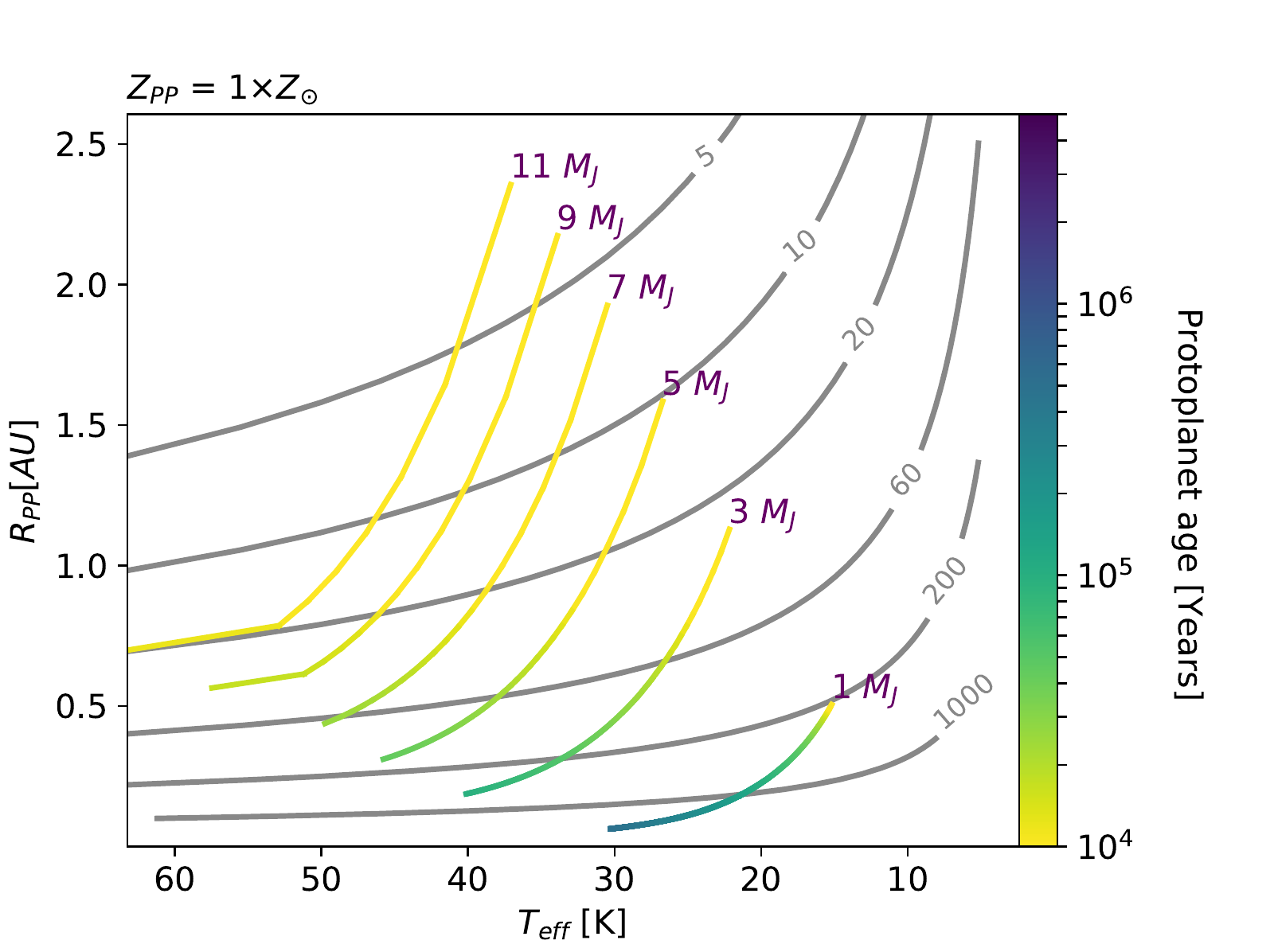}
\includegraphics[width=1.0\columnwidth]{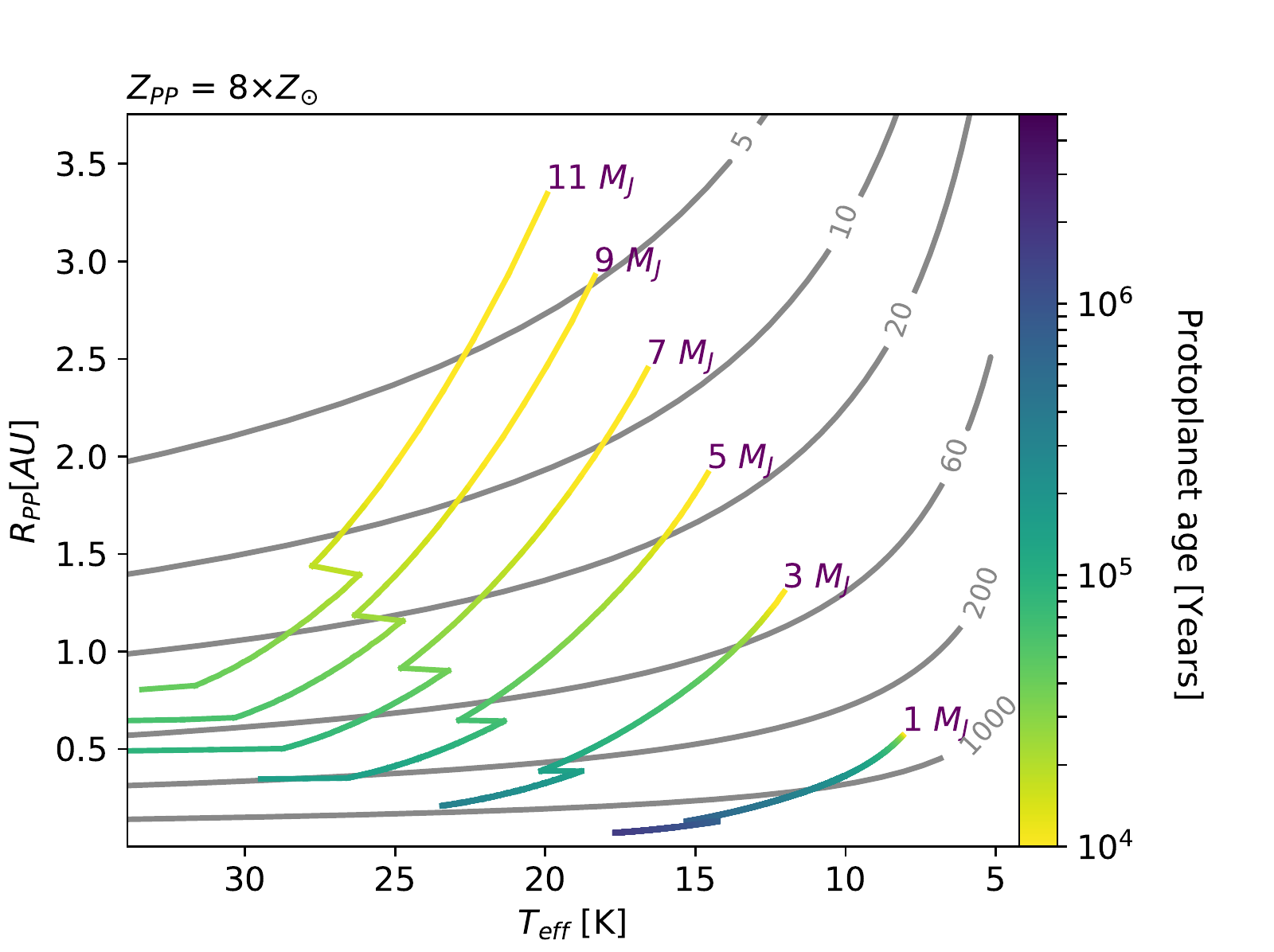}
\caption{Protoplanet radius ($R_{\rm PP}$) against effective temperature ($T_{\rm eff}$) for different planet masses. Left: $Z_{PP}=Z_{\odot} = 0.015$, right: $Z_{PP}=8 \times Z_{\odot}=0.12$. The evolution track lines are coloured by protoplanet age from yellow to dark blue. The grey lines mark contours of equal required observation time in minutes. Note that lower mass, older and more metal-rich planets are the hardest to detect.}
\label{fig:evo_tracks}
\end{figure*}

So far we have assumed that the surface temperature of a protoplanet is set by irradiation from its host star and tested just two representative values for this value. To make further progress we now combine our synthetic observation results with a self-luminous protoplanet evolution model from \cite{Nayakshin15c} in order to calculate the necessary integration time as a function of protoplanet mass and age. These results are independent of shadowing by the disc, so require fewer free parameters to model. We shall discuss the distinction between self-luminous and irradiation-dominated planets further in section \ref{sec:IMS}.

The \cite{Nayakshin15c} model assumes an isentropic profile inside the protoplanet, with a convective interior and radiation through a grey atmosphere at the surface, using the \cite{ZhuEtal12a} fits to dust opacity. The model is detailed fully in \cite{Nayakshin15c} and was used to study metallicity trends in giant planets in \cite{Nayakshin15d} and \cite{NayakshinFletcher15} as well as to study feedback from protoplanet cores in \cite{Nayakshin16a}. \cite{Nayakshin20-TW-Hya} have recently compared the evolutionary tracks for planets computed with this code with those computed with the more detailed stellar evolution approach \citep{VazanHelled12} at the same dust opacity. An agreement sufficiently close for the purposes of our study was found.

The model provides radius and temperature information during the course of a protoplanet's evolution. Meanwhile, our synthetic observations show that it takes $\sim$ 10 minutes to observe a  1.5 AU radius, 30 K protoplanet. By treating the protoplanet as a black body emitter, we can take this fiducial result and scale it in order to obtain the required observation time to achieve detection  (SNR = 5) of an arbitrary protoplanet in terms of $R_{\rm PP}$ and $T_{\rm eff}$:

\begin{equation}
    O_{\rm time} = 10 \times \dfrac{B_\nu(30 \rm K)}{B_\nu(T_{\rm eff})}  \left( \dfrac{1.5 \rm AU}{R_{\rm PP}} \right)^2 [\rm minutes]\;,
    \label{eq:obs_time}
\end{equation}
where $B_\nu$ is the Planck function for spectral radiance. This formula reproduces the SNR values obtained for the other combinations of radius and surface temperature used in Section \ref{sec:results}.
We can now make estimates for the required detection time for any protoplanet over its lifetime. 

The final parameter that we introduce into the models is the metallicity of the protoplanet. As discussed already, a combination of enrichment at birth and subsequent efficient pebble accretion can substantially enhance the metallic compositon of young protoplanets \citep{MillerFortney11, HumphriesNayakshin18}.
To reflect this we examine two extreme cases, one for protoplanets of solar metallicity ($Z$=0.015) and the other for metallicities of eight times solar ($Z$=0.12), assuming that dust opacity in the protoplanet is directly proportional to its bulk metallicity $Z$. 
By $10^5$ years, all of these models protoplanets have reached a relatively `compact' state with sub-AU radii, this already implies that detection after this age will require at least 100 minutes of observation time. See Appendix \ref{app:models} for an overview of the \cite{Nayakshin15c} models at both metallicities.

In Figure \ref{fig:age_obs} we visualise the lifespan of protoplanets of varying mass, coloured by the required ALMA observation time in minutes. 
The left panel shows protoplanets with solar metallicity while on the right they have the enhanced value of eight times solar.
In the solar metallicity case, we see that protoplanets more massive than $7 M_J$ are luminous enough that they may be observed in under 10 minutes by ALMA. However, they only survive in this state for at most $10^5$ years and so the chance to observe them before they collapse is very low. The planet forming associations at 140 parsecs, Taurus and Ophiuchus, are at least $10^6$ years old \citep{SoderblomEtal14}, although there is some spread in individual system ages.
In contrast we see that a $1 M_J$ protoplanet may survive for $2 \times 10^ 6$ years, but has a necessary integration time of over 1000 minutes.

For the high metallicity models on the right, protoplanet lifetimes are extended by a factor of $\sim$ 5 at all masses. These protoplanets are more likely to survive to be observed in the local associations, but are correspondingly dimmer and so harder to detect. Only protoplanets younger then $5 \times 10^4$ years can be detected in under 20 minutes of observation time. 

In Figure \ref{fig:evo_tracks} we visualise the relationship between protoplanet radius and effective temperature, again with $Z=Z_{\odot}$ on the left and $Z = 8 \times Z_{\odot}$ on the right. The plotted lines mark the evolution tracks for protoplanets of different masses, these are coloured by the corresponding age of the protoplanet from yellow to dark blue, as shown in the colourbar. In this way, these figures are analogous to stellar HR diagrams. The grey lines mark contours of required observation time in minutes, we can see that longer detection times are required to detect older protoplanets of any mass. Newly born protoplanets contract and dim rapidly at first, before their evolution slows as they grow older. The required detection time increases by around a factor of ten over the lifetime of a typical protoplanet in these figures. Note, the abrupt changes in planet luminosity seen in the high metallicity models occur at dust opacity transitions when certain species vaporise \citep[see][and Figure \ref{fig:PP_models} for more detail.]{ZhuEtal12a}.

The results in this section show that low mass and/or high metallicity protoplanets require at least an hour of ALMA integration time. They are unlikely to be detected serendipitously in local association disc surveys, instead they will require dedicated observations with at least a few hundred minutes per source \citep[e.g., see][]{TsukagoshiEtal19}.

\section{Discussion}

\label{sec:discussion}
\subsection{Main results}

In this paper we aimed to quantify the chance of detecting a first-core-like, GI pre-collapse protoplanet orbiting a protostar with an age of a few million years, using ALMA observations. Our results demonstrate that observing long-lived GI protoplanets will typically require hundreds of minutes of ALMA observation time per source, since protoplanets are generally dim and compact (sub AU radius). They are highly unlikely to be spotted in surveys of the `one minute per source' type such as \cite{AnsdellEtal17, LongEtal18} or \cite{CiezaEtal19}. Dedicated observational campaigns such as DSHARP with hour long on source times stand a better chance of detecting protoplanets, though  surveys with a few hundreds of minutes per source are necessary to detect protoplanets older than $10^5$ years.

In general, there exists a tension between protoplanet luminosity and the duration of the pre-collapse phase. High mass protoplanets are very bright and so could be readily detected by ALMA, but their lifetimes are short, making them unlikely to survive even to the class I disc phase. 
For example, at solar metallicity ($Z = Z_{\odot}$) only protoplanets less massive than $5 M_J$ survive for longer than $10^5$ years, while only Jupiter mass protoplanets reach ages above $10^6$ years. The trade off is that these low mass protoplanets are less luminous. From Figures \ref{fig:age_obs} and \ref{fig:evo_tracks} we can see that detecting a $5 M_J$ protoplanet at $10^5$ years requires 100 minutes of Band 6 integration time, while detecting a 1 $M_J$ protoplanet at $10^5$ years requires 500 minutes, rising to more than 1000 minutes at $10^6$ years.

This tension is shifted for protoplanets with higher bulk metallicities. From Figures \ref{fig:age_obs} and \ref{fig:evo_tracks} we can see that while high metallicity protoplanets ($Z = 8Z_\odot$) live longer (by around a factor of five in lifetime), they are less luminous and so making a detection will require a comparatively longer integration time. Considering that the nearby associations at 140 parsecs have typical ages of a few million years \citep[or more][]{SoderblomEtal14}, the extended lifetimes of metal enriched protoplanets means that they have a higher possibility of being detected in these systems.

Given these considerations, we find that a high metallicity protoplanet with a mass of $\sim$ 2-4 $M_J$ is the best case detection candidate for observations made in a nearby association with ALMA. Utilising ALMA configurations with smaller beam sizes increases the chance that a given protoplanet could be resolved from the outer disc edge. Applying the gap opening relations found empirically by \cite{LodatoEtal19} to disc truncation, we showed in Figure \ref{fig:PP_lengthscale} that a maximum beam size of 15 AU is necessary in order to resolve Jupiter mass protoplanets orbiting outside of circumstellar discs. 
The chance that a protoplanet would be obscured by an optically thick disc may also be reduced by observing discs that are close to face-on, though we do not quantify this fraction here.

\subsection{Pathways to a protoplanet detection}\label{sec:enhancing}

Ultimately, the occurrence rate of GI, and therefore the number of expected protoplanets, is currently unknown. Disc fragmentation may conceivably take place in every embedded system with an infall rate high enough to continually increase the disc mass \citep{Clarke09}, but could also be markedly rarer \citep[e.g. ][]{KratterEtal10,ViganEtal17}. 
Motivated by our previous research that suggested that GI could occur in as many as tens of percent of systems \citep{HumphriesEtal19}, in this paper we hoped to investigate the feasibility of detecting a protoplanet around a `typical' $10^6$ year old disc observed by ALMA. GI discs evolve considerably over their lifetimes, and so such a disc may have been massive enough to be gravitationally unstable during its youth. Detecting a protoplanet around a compact, million year old disc would provide evidence for a high occurrence frequency for GI.

However, we found that detecting protoplanets after $10^6$ years requires an unreasonably long observation time (above 500 minutes). Order Jupiter mass protoplanets may well exist around small discs at a few $10^6$ years, but a large scale survey approach will fail to detect such dim objects unless the integration time is several hundred minutes per source. Instead of indiscriminate surveying, we suggest that the best chance of making a protoplanet detection is to target systems that have the highest chance of hosting one.

\subsubsection{The youngest systems}\label{sec:youngest}

An immediate tactic is to focus observations on the young ($\sim 10^5$ year old) and massive discs that are most likely to have experienced GI during birth. As well as being more likely to host protoplanets, protoplanets in young systems are more luminous and so require shorter integration times. The time required to detect a typical protoplanet increases by a factor of ten over their lifetime. For example, Figure \ref{fig:evo_tracks} shows that it takes 100 minutes to detect a high metallicity $3 M_J$ protoplanet at $10^5$ years, but this value rises to 1000 minutes at $10^6$ years. Furthermore, the more massive protoplanets typically collapse before $10^5$ years, observing the youngest systems therefore increases the mass range of protoplanets that are able to be detected.

Signs of disc instability will still be visible up to ages of a few $10^4$ years \citep{HallEtal19}, and so work that self-consistently considers protoplanets born in fragmenting disc simulations \citep[such as][and others mentioned in the introduction]{ZakhozhatEtal13} will be very relevant in this case.
Although such young systems may well be embedded in a natal core or cloud, we do not anticipate this to be an issue since the foreground material will be optically thin at mm wavelengths, as in the case of HL Tau \citep{Alma2015}.

\subsubsection{Intermediate mass host stars}\label{sec:IMS}
Many studies now imply that GI operates most readily around higher mass stars. For example, the occurrence rate of giant planet companions has been observed to increase as host mass increases \citep{NielsonEtal19}. Analytic arguments based on a Toomre $Q = 1$ disc, summarised by \cite{KratterL16} also suggest the critical disc-to-star mass ratio ($M_d/M_*$) for a gravitationally stable disc scales with the stellar mass as
\begin{equation}
    \left(\frac{M_d}{M_*}\right)_{crit} 
    \propto T^{1/2}r^{1/2}M_*^{-1/2},
    \label{equn:KratterLodato}
\end{equation}
where $T$ is the temperature at radius $r$ in the disc. This shows that discs around more massive systems are more likely to undergo fragmentation and so more likely to produce protoplanets. The companion papers \cite{HaworthEtal20} and \cite{CadmanEtal20} recently extended this result by including stellar irradiation in their pseudo-viscous models and SPH simulations. They found that irradiation from low mass stars may increase the critical mass threshold for disc fragmentation above the previously calculated values, increasing the chances that protoplanets will have been formed around more massive stars.

Furthermore, Equation \ref{eq:obs_time} and Figure \ref{fig:age_obs} show that the hotter the effective temperature of the planet, the shorter the integration time needed to detect it. 
Since more massive stars are more luminous, irradiation from these sources may increase the surface temperature of nearby protoplanets above their self-luminous values. In this `irradiated protoplanet' regime, the surface temperature is very nearly the irradiation temperature, and the required time for a detection decreases as a function of the stellar luminosity. 
Re-purposing Equation \ref{eq:obs_time} for our $0.25 L_{\odot}$ star, we find

\begin{equation}
        O_{\rm time} = 250 \;
    \frac{B_{\nu}(30 K)}{B_{\nu}(T_{irr})} \left( \dfrac{0.3\; \rm AU}{R_{\rm PP}} \right)^2 [\rm minutes],
    \label{eq:obs_time_irr}
\end{equation}{}

with $T_{irr} = (L_* /(4 \pi r^2 \sigma_{SB})^{1/4}$ and where we have used $R_{\rm PP} = 0.3$~AU as more appropriate for an older protoplanet. By evaluating the spectral radiance function, Equation \ref{eq:obs_time_irr} shows that a two hour long observation of a source 140 pc away could discover planets larger than $0.3$~AU in radius if they have surface temperatures of 60 K. Such an irradiation temperature profile may be found at 40 AU around a star with a luminosity of $4 L_{\odot}$, not an unlikely occurrence since young stars are typically hotter than their main sequence counterparts. 

Detecting a GI pre-collapse protoplanet with ALMA is therefore more likely around more massive protostars simply because they are more likely to be brighter. Orbital separation also becomes important in this irradiation dominated regime, adding weight to the requirement for detection surveys to be able to resolve protoplanets close to the disc edge.
Increased surface temperatures due to irradiation suggests that going after more massive, e.g., Solar, and higher mass, systems such as $M_*\sim 2 M_\odot$ Herbig Ae stars, represents a good choice for a protoplanet detection campaign.

\subsubsection{Targeting massive non-accreting planet candidates}
\label{sec:CO}

ALMA is capable of uncovering massive planet candidates in young protoplanetary discs via the indirect signatures that embedded planets have on the host discs: annular gaps and rings \citep{BroganEtal15,HuangEtal16,LongEtal18,DSHARP7}, spiral density features \citep{DongEtal18_MWC758}, and CO velocity kinks \citep{TeagueEtal18, PinteEtal18, CasassusPerez19}.

\cite{PinteEtal20} analysed ALMA CO data for the 18 circumstellar discs observed by the DSHARP program \citep{DSHARP1}. They found that 8 of the discs show velocity kinks (one system hosts two), corresponding to planets with masses $\sim (1-3)\mj$\footnote{Though these calculated masses are between 4 to 10 times higher than the masses predicted by \cite{LodatoEtal19} using the results of dust gap opening models.}. The strongest indirect evidence that these kinks are made by planets is that all of them are either localised within a dust gap or lie at the end of a spiral arm. Tellingly, none of these candidate planets have yet been detected via gas accretion signatures such as H$\alpha$ emission. The lack of H$\alpha$ emission is a unique identifier of pre-collapse GI protoplanets, in stark contrast to post-collapse planets or planets made by Core Accretion. We can see this by considering that the shock temperature is comparable to the virial temperature at the surface of the planet,

\begin{equation}
    T_{\rm sh} \sim \frac{GM_{\rm p}\mu}{k_{\rm B} R_{\rm p}} \sim 
    \begin{cases} 
    700 \textrm{ K} &\mbox{if } R_{\rm p} = 0.3 \textrm{ AU} \\ 
2 \times 10^5 \textrm{ K} & \mbox{if } R_{\rm p} = 2 \; R_{\rm J} \end{cases}
    \label{Tshock}
\end{equation}

The first line in this equation is for pre-collapse GI protoplanets where we set $R_{\rm p}\sim 0.3$~AU, and the second for post-collapse planets. To emit the H$\alpha$ line, the shocked gas temperature must exceed $\sim 10^4$~K. We can see that this is easily satisfied by the latter but not by the former.
A lack of H$\alpha$ line emissions is therefore consistent with the scenario in which the \cite{PinteEtal20} planets are the dusty pre-collapse protoplanets that we study here. These candidates may therefore be promising targets for the direct detection of a protoplanet in ALMA dust continuum observations.

\subsubsection{Dust-losing planets}\label{sec:dust-losing}

Recently, \cite{TsukagoshiEtal19} found a highly statistically significant excess emission from a dust `clump' of an $\sim$ AU scale, embedded at 52 AU in the outer disc of a well known $10^7$ year old system TW Hya. Modelling the emission as optically thin, the authors deduced a dust mass of $\sim 0.03\mearth$, although they note that there may be much more dust if the source is optically thick (see also Section \ref{sec:o-thick}). Modelling of this ALMA observation by \cite{Nayakshin20-TW-Hya} showed that the excess emission may be due to a dust outflow from a disintegrating protoplanet. As the outflow leaves, its surface area exceeds that of the protoplanet by a large factor. This shows that a dust outflow from a planet may in fact be much more readily detectable than the planet itself. Potential dust outflows from disrupting protoplanets could be further followed up with ALMA; observing them in longer wavelengths may allow ALMA to peer deeper through these outflows to discern the underlying protoplanet.

\subsubsection{Misidentification of optically thin features}\label{sec:o-thick}

The \cite{Nayakshin20-TW-Hya} analysis in the previous section showed that interpreting a GI protoplanet as an optically thin feature would lead to a dramatic underestimate in its mass. For example, we can calculate an `optically thin dust mass' for the $3 M_J$ protoplanet from the high resolution right hand panel of Figure \ref{fig:surveys} using the continuum flux density \citep[eg.][]{BeckwithEtal90}

\begin{equation}
    M_{\rm dust} = \dfrac{F_{\nu} d^2}{\kappa_{\nu} B(T)}.
    \label{eq:dust_mass}
\end{equation}

$F_{\nu}$ is the total flux of the protoplanet at 1.3 mm, d is the distance (140 pc) and $B(T)$ is the Plank Function. We follow \cite{TsukagoshiEtal19} and take the optical depth $\kappa_{\nu}$ at 1.3 mm to be 2.3 cm$^{2}$g$^{-1}$. A flux of 0.32 mJy, a distance of 140 pc and a temperature of 30 K yields a total dust mass of 0.1 $M_{\oplus}$, almost 10,000 times less than the total gas and dust mass of our synthetic protoplanet. 
Increasing awareness of the observational signatures of protoplanets will help to characterise accidental observations and help to identify systems that warrant follow-up observations.

\subsubsection{Individual targets}\label{sec:individual}

Finally, we speculate on individual protostars for which, in our opinion, observed features may indicate the presence of a protoplanet.

The PDS 70 system hosts two confirmed gas giant planets PDS 70b and PDS70c, orbiting the host star at $\sim 20$~AU and 35 AU \citep{IsellaEtal19} and detected in H$\alpha$ line, thus both accreting gas. The planets, if formed via GI, are in the post-collapse configuration. However, in addition to these planets, \cite{KepplerEtal19} suggests that may be a massive planet at $\sim 70$~AU, where the dust emission of the main ring in this system peaks. One of the sides of the rings is brighter in the ALMA continuum \citep{LongEtal18}. Further, the same side of the ring hosts a potential point source detected in the CO continuum. What makes this point source additionally interesting is that it shows a velocity offset from the local Keplerian rotation of about 1 km/sec. Preliminary results show that outflows of this velocity magnitude from pre-collapse protoplanets are expected (Nayakshin et al., in prep.).

HD 163296 is another interesting candidate target. \cite{DSHARP9_HD163296} find that ALMA emission from this source shows ringed but significantly azimuthally asymmetric structure, which may perhaps be indicative of an embedded protoplanet in one of the rings. 

Finally, \cite{CasassusPerez19} detected a velocity flip in the disc of the Herbig Ae star HD100546 at a separation of $\sim 27$~AU, consistent with a planet of a few Jupiter masses. As with the potential source discussed in PDS 70 by \cite{KepplerEtal19}, the location of the source coincides with a ridge inside the dust ring, which is opposite to the usual expectation that planets are found within dust gaps. This may indicate that the dust in the rings surrounding the potential protoplanet may originate from its outflow, which naturally explains the protoplanets spatial coincidence along the ridge \citep[see for example][]{Nayakshin20-TW-Hya}.

\subsection{The significance of a protoplanet detection}

Detecting even a single protoplanet would have a significant impact on the field of planet formation. Given the challenges presented in this paper it is likely that a much larger, undetected population may also exist. It is also important to understand why no such detections have been made to date; is this because they are rare or simply difficult to observe. It is also quite possible that both of these statements are true.

To start with, detecting a  first-core like protoplanet with a mass less than 10 $\mj$ would demonstrate that GI is able to form planetary mass objects in addition to brown dwarfs and stellar companions. This is important since the minimum mass for a GI fragment is still disputed \citep{KratterL16}. If GI can form objects in the planetary mass regime, it will help to explain wide-orbit gas giant systems such as HR 8799 \citep{MaroisEtal08,MaroisEtal10}, or the presence of multiple gas giants around the M dwarf GJ 3512 \citep{MoralesEtal19}.

There is currently disagreement concerning the relative frequency of GI as a planet formation mechanism compared to the alternative core accretion theory.
The current, widely accepted figure is that GI occurs in just 2 \% of systems, motivated by a lack of detected wide-orbit gas giant planets in Direct Imaging (DI) surveys \citep{ViganEtal17}. However, gap opening in these models may be too efficient \citep{MalikEtal15,MullerEtal18} and gas accretion is neglected entirely, potentially leading to over-predictions for the number of surviving wide-orbit gas giants and brown dwarfs. For instance, \cite{HumphriesEtal19} recently presented population synthesis simulations which showed that the DI observations could also be consistent with GI occurring in as many of tens of percent of systems.

Meanwhile in the inner disc, trends in gas giant occurrence frequency with stellar metallicity indicate that there is a turnover mass in the range 0.5-25 $\mj$ between the dominance of CA vs GI \citep{SantosEtal17, Schlaufman18, MaldonadoEtal19}, though the exact value remains uncertain. \cite{HumphriesEtal19} showed that planet formation via disc fragmentation must occur in at least tens of percent of systems if GI is required to explain a significant fraction of these sub-AU massive gas giants, which is at odds with the 2\% result for GI frequency. \cite{SuzukiEtal18} also appealed to GI in order to explain massive gas giants after finding that gas accretion may be too efficient in the current generation of CA models. A protoplanet detection would help to confirm that GI can indeed form gas giants and help to resolve the cause of the turnover mass in the occurrence frequency of gas giants with system metallicity.

Detecting a long-lived protoplanet would require some explanation for why protoplanets exist at wide orbits while evolved gas giants do not. Of course, there is no guarantee that all of the mass in a protoplanet will collapse into the `second core', for instance \cite{GalvagniEtal12} used 3d simulations to show that protoplanets typically lost half of their mass during their rapid collapse down to a gas giant planet. Furthermore, protoplanets may also be disrupted by internal feedback if they form a 20-30 $M_{\oplus}$ core before they undergo Hydrogen dissociation collapse \citep{Nayakshin16a,HumphriesNayakshin19}.
This mechanism depletes the number of wide orbit protoplanets as well as generating a population of released rocky cores and debris at wide orbits \citep{NayakshinCha13}, solving two of the current weaknesses of GI as a planet formation mechanism.

\section{Conclusions}
\label{sec:conclusions}

In conclusion, we find that it is possible to detect pre-collapse wide-orbit protoplanets with ALMA and that such a detection would have important implications for planet formation and protoplanetary disc evolution. Searches for GI protoplanets with ALMA rather than the more traditional direct imaging techniques in the optical or near infrared \citep[e.g.,][]{ViganEtal17} are important because pre-collapse GI planets may not be bright in these higher frequencies and hence hide ``in the plain sight". Our model is not perfect, as self-consistently modelling these systems on very long timescales is computationally unfeasible. What we suggest is that this toy model is broadly consistent with what is expected from GI, and detection of these objects, or consistent absence of detection, will constrain the occurrence of planet formation through the GI pathway.

\smallskip

We have shown that ALMA has the capacity to detect long-lived protoplanets around discs that underwent gravitational instability during their formation, though configurations with projected beam sizes of less than 15 AU should be used in order to resolve these objects from the disc.

We used radiative transfer simulations to show that at least several hundred minutes of ALMA integration time will be necessary in order to detect a protoplanet with a SNR greater than 5, and the smaller beam sizes are favoured. 





Protoplanets generally have sub-AU radii and so order AU beam sizes will not decrease the sensitivity of a detection. We expect that a protoplanet detected in this case would have a mass of $3-5 M_J$, any smaller and it will be impractically dim, any larger and it's pre-collapse phase is too short (less than $10^5$ years). The necessary integration time in this case will be at least 100 minutes, though could be as high as 500 minutes. High metallicity protoplanets will be longer lived, but dimmer. Finally, we expect that such a protoplanet will likely be external to any surviving disc, since migration and gas accretion of protoplanets embedded in discs reduce their chance to survive to the million year mark.
High resolution surveys such as DSHARP provide the best chance of making such a detection, provided their integration times per source are increased to a few hundred minutes.

We note that this work has focused on dust as the tracer of protoplanets, but considerable work on the chemical evolution of these bodies has also been conducted \citep[see, e.g.][]{IleeEtal17,quenardetal2018}. In general, it has been realised that molecular line observations of protoplanetary accretion discs reveal more information about the object than continuum observations alone, such as the precise location of a hiding core accretion planet \citep{TeagueEtal18,PinteEtal18,pinteetal2019, PinteEtal20},  and characteristic signatures of graviatational instability \citep{halletal2020}, and so this is an important avenue for future work.

\cite{HallEtal19} showed that it is necessary to target very young discs in order to directly detect the characteristic spiral features of GI, something that is very challenging due to the embedded nature of these systems. Protoplanets act as leftover `by-products' of earlier GI fragmentation, detecting one would provide valuable information about earlier episodes of gravitational instability in accretion discs. Such a detection would also constrain the relative rates of GI planet formation, the mass spectrum of GI fragments and provide clues about the early evolutionary stages of million year old discs. For these reasons we believe that observing a pre-collapse protoplanet would have a transformative effect on the field of planet formation and we encourage future efforts from the community to detect these objects.

\section*{Acknowledgements}
JH and SN acknowledge support from STFC grants ST/N504117/1 and ST/N000757/1. TJH is funded by a Royal Society Dorothy Hodgkin Fellowship. CH is a Winton Fellow and this work has been supported by Winton Philanthropies/The David and Claudia Harding Foundation.

\section*{Data Availability}
The data underlying this article will be shared on reasonable request to the corresponding author.

\bibliographystyle{mnras}
\bibliography{humphries2}


\appendix

\section{RADMC-3D convergence testing}
\label{app:conv}
\begin{figure}
\includegraphics[width=1.0\columnwidth]{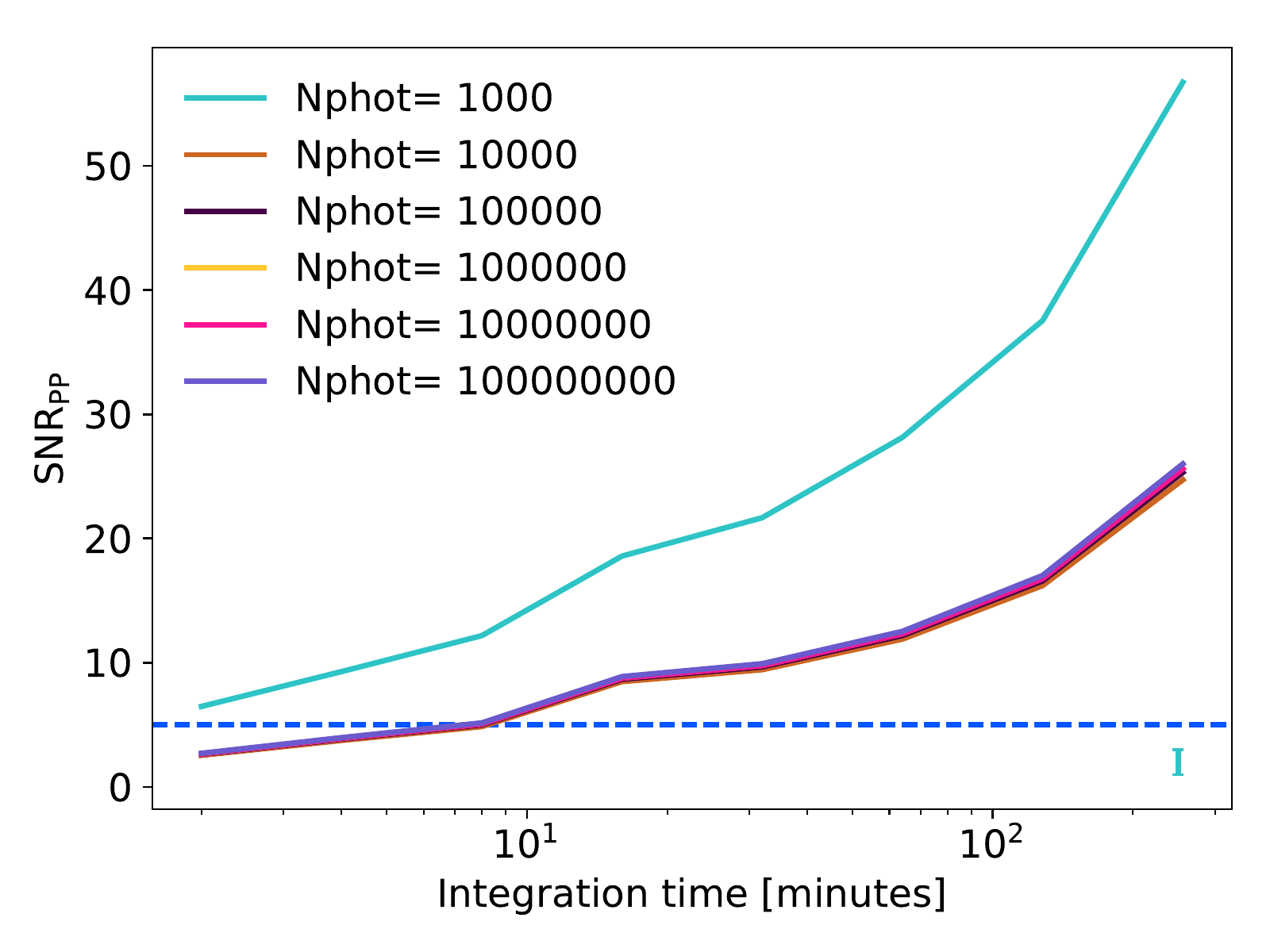}
\caption{Convergence testing with RADMC-3D. SNR ratio for a variety of different photon packets. Due to the simple geometric nature of the problem the SNR converges after only $10^4$ packets.}
\label{fig:conv}
\end{figure}

Due to the simple geometric nature of the problem the solution converges with only 10$^4$ photons, \textbf{as shown in Figure \ref{fig:conv}}. There is only a $\sim$ 5\% difference in the SNR for $10^4$ and $10^8$ photons. In order to produce smooth images, we have used $10^7$ for most of the runs in this paper and so we are satisfied that the SNR values are fully converged.

\section{Concatenated observations}
\label{app:concat}

\begin{figure}
\includegraphics[width=1.0\columnwidth]{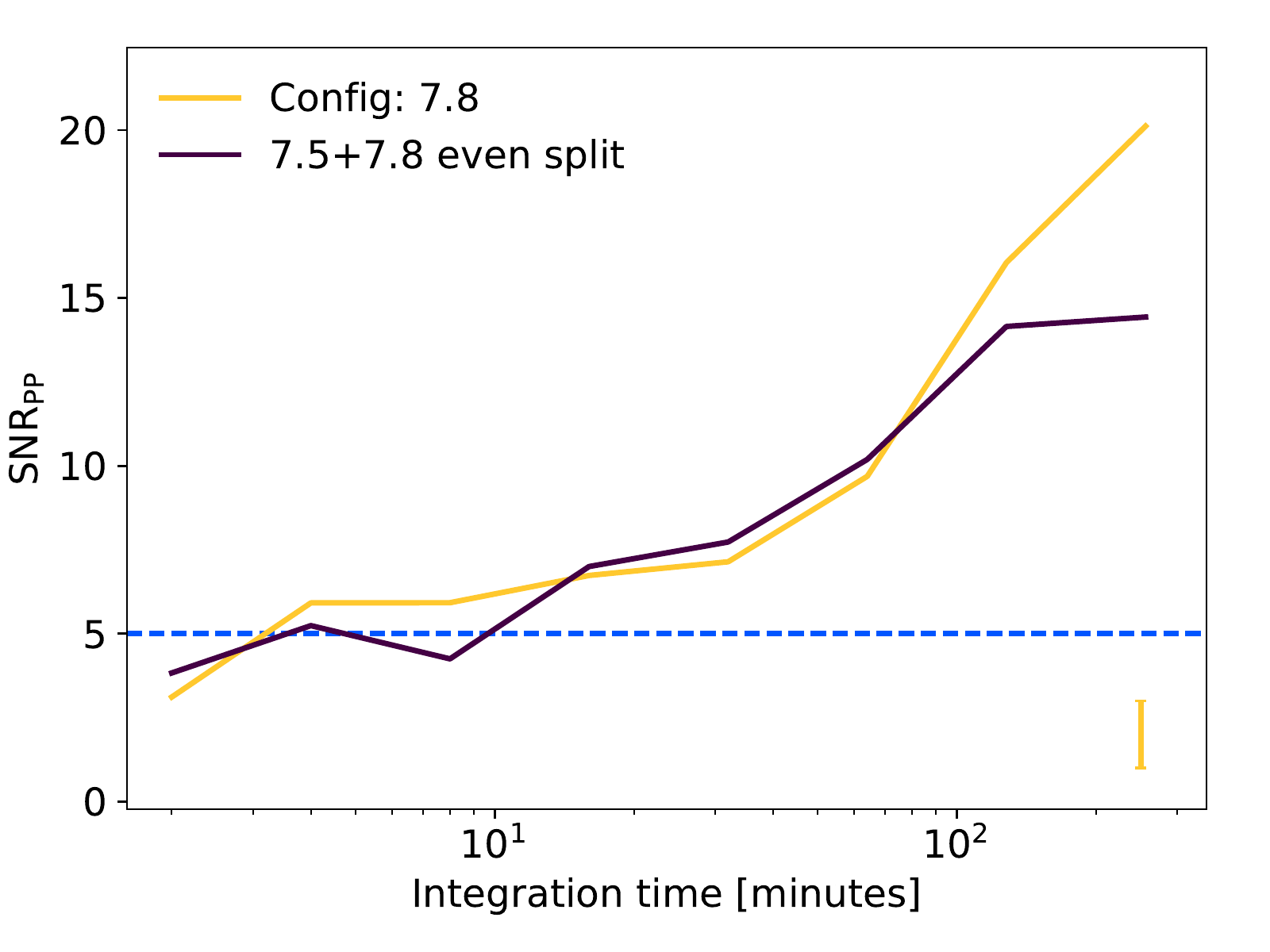}
\caption{Signal to noise value for protoplanet detections, comparing observations made with a pure 7.8 configuration (blue) to a 50:50 split of 7.5 and 7.8 configurations (pink). Concatenating measurement sets will not increase the SNR value for protoplanets smaller than the smallest beam size.}
\label{fig:SNR_concat}
\end{figure}

It is common practice to concatenate multiple ALMA observations in order to obtain higher sensitivity images with lower total observation times. In Figure \ref{fig:SNR_concat} we show that this is not beneficial for protoplanet detections, since protoplanets will typically be smaller than the beam size scale. The SNR value for the pure 7.8 configuration agrees with the value from a 50:50 mix of 7.5 and 7.8 configurations, the one stand deviation errorbar is shown in the bottom right of the figure. For this reason we find that concatenating measurement sets will not increase the chance of detecting long-lived, wide-orbit protoplanets.

\section{Integration times at 100 AU}
\label{app:int100}

\begin{figure*}
\begin{tabular}{ccc}
    \includegraphics[width=0.66\columnwidth]{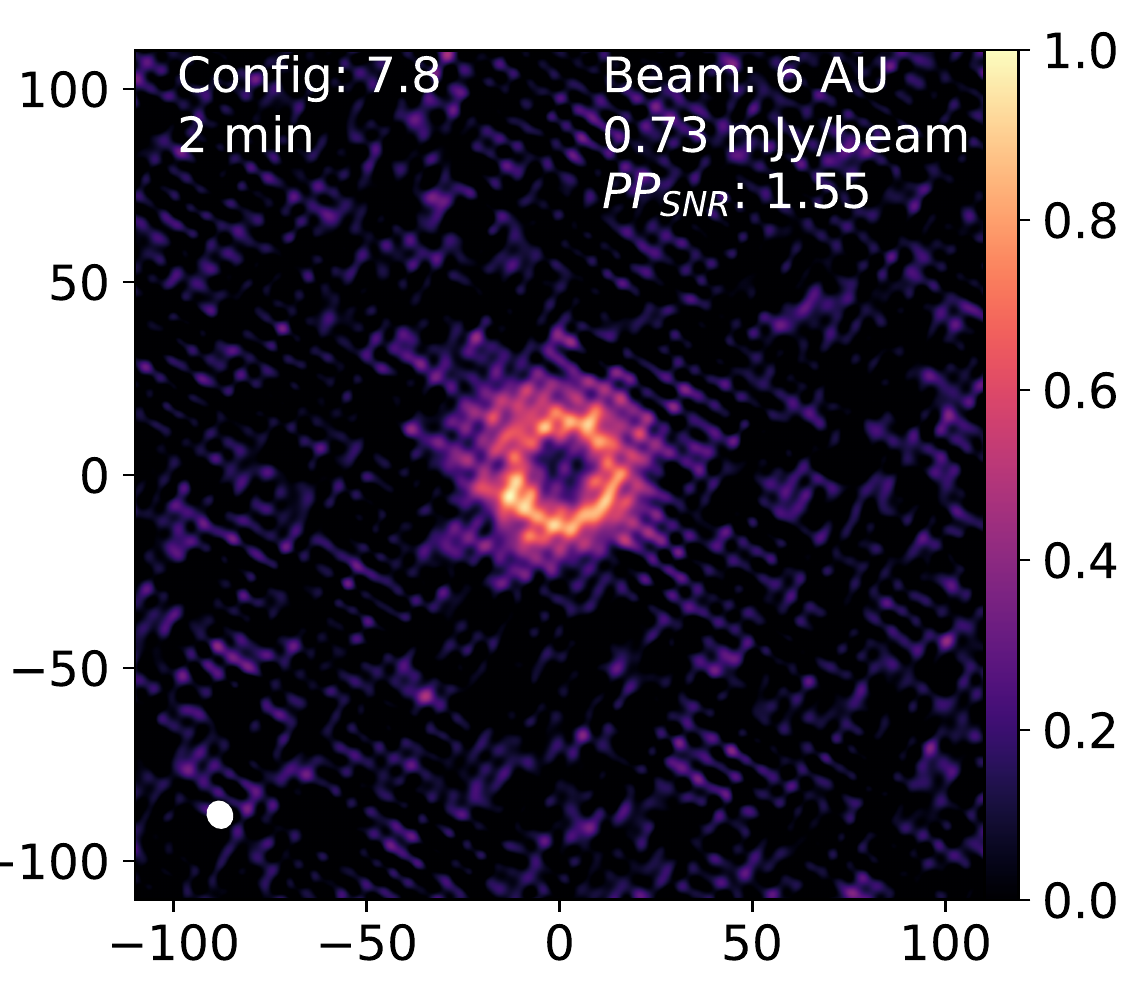} &
    \includegraphics[width=0.66\columnwidth]{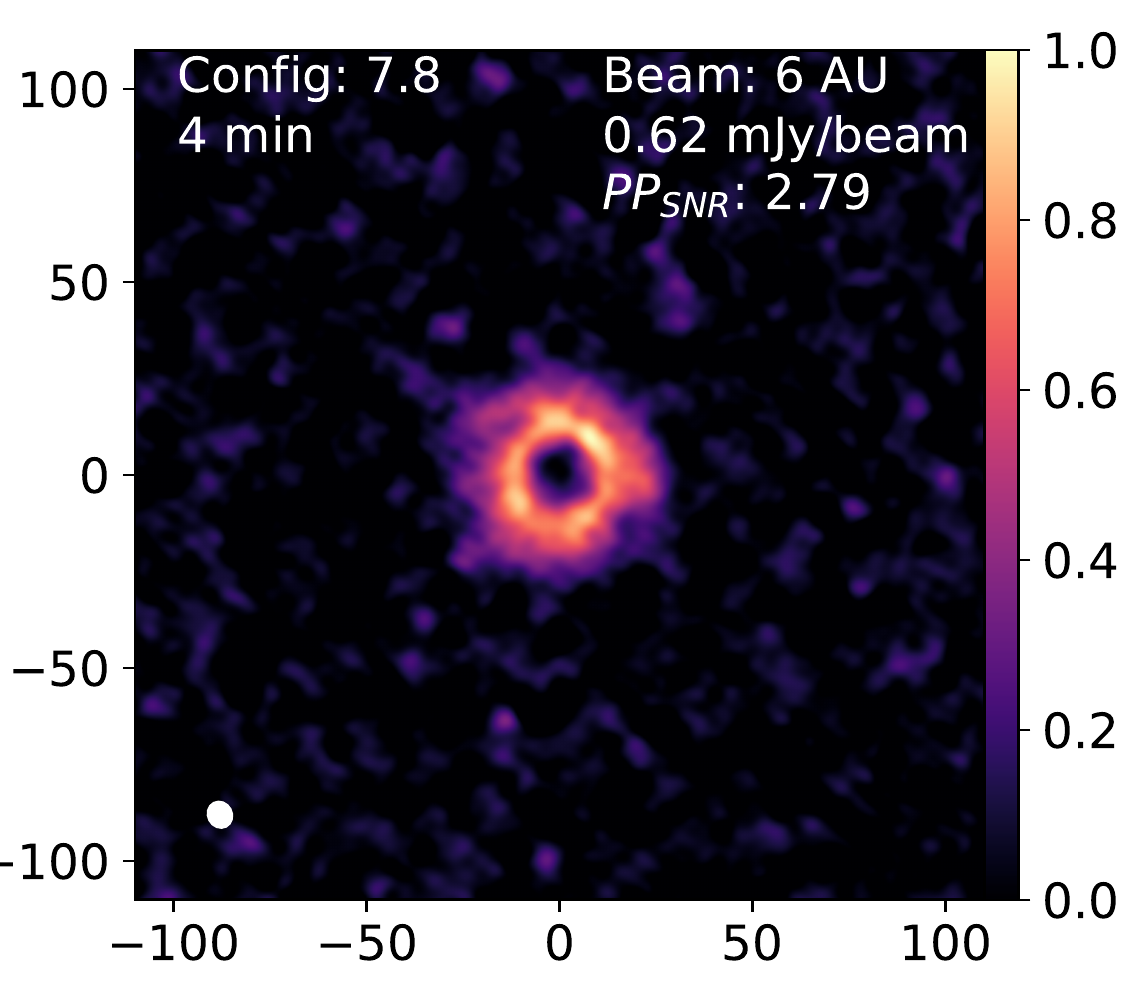} &
    \includegraphics[width=0.66\columnwidth]{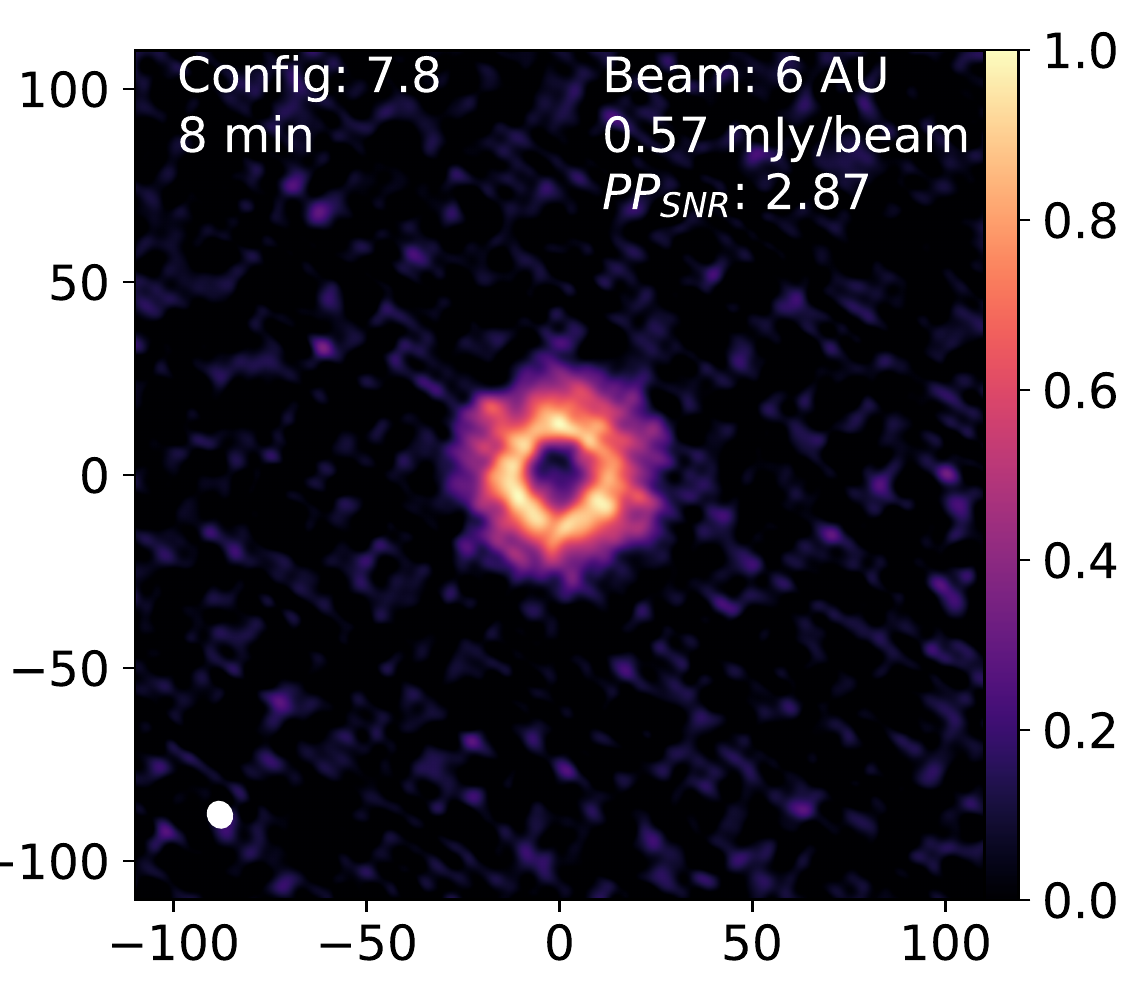}\\
    a) & b) & c) \\    
    \includegraphics[width=0.66\columnwidth]{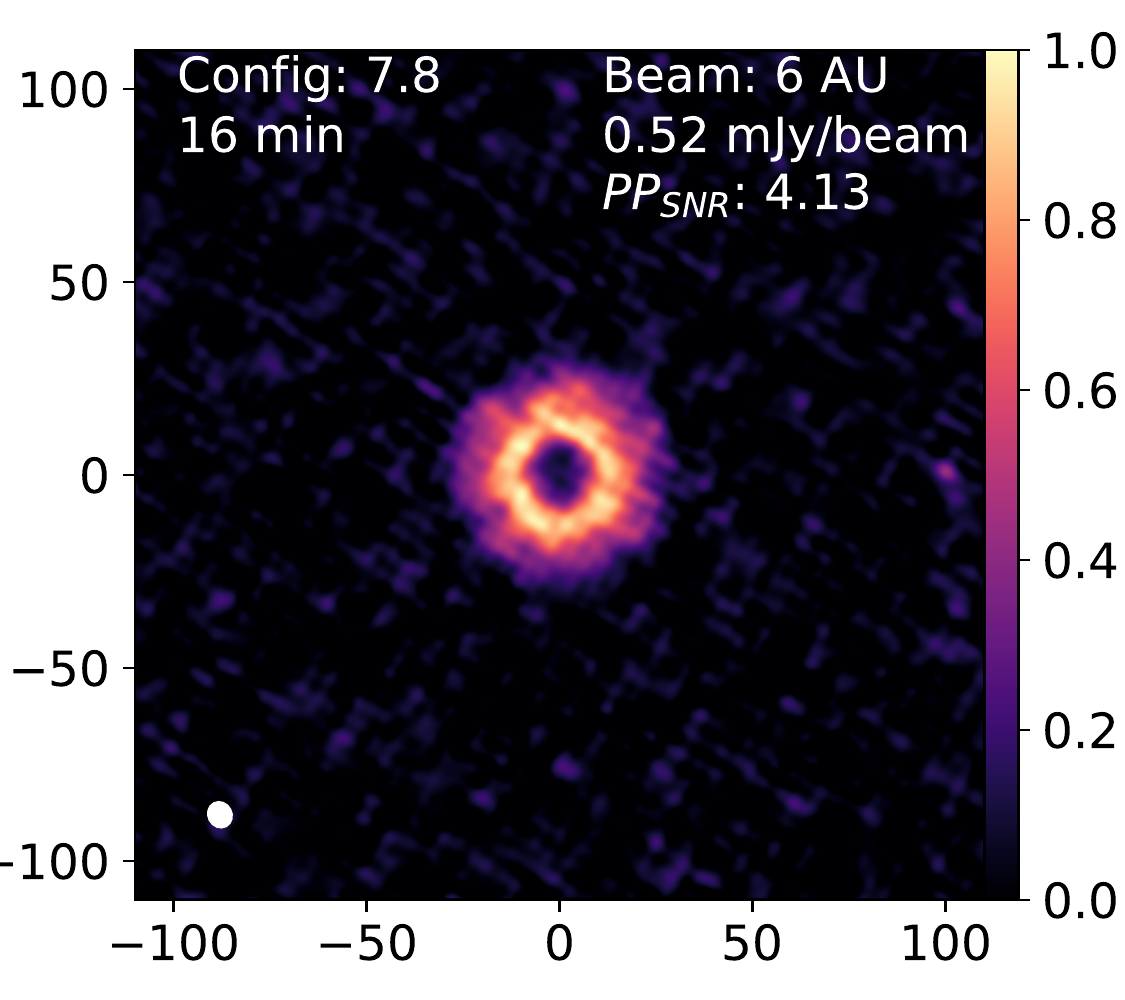} &
    \includegraphics[width=0.66\columnwidth]{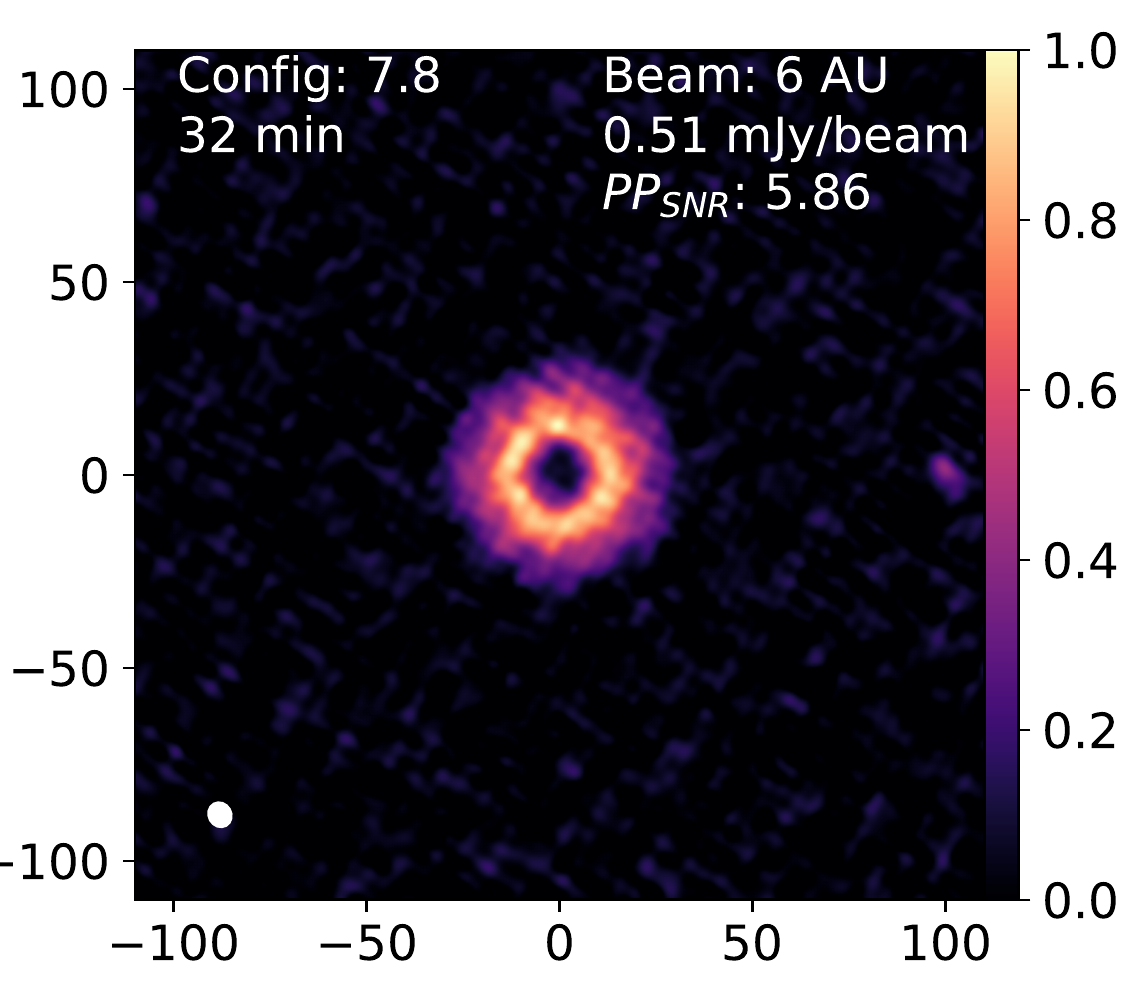} &
    \includegraphics[width=0.66\columnwidth]{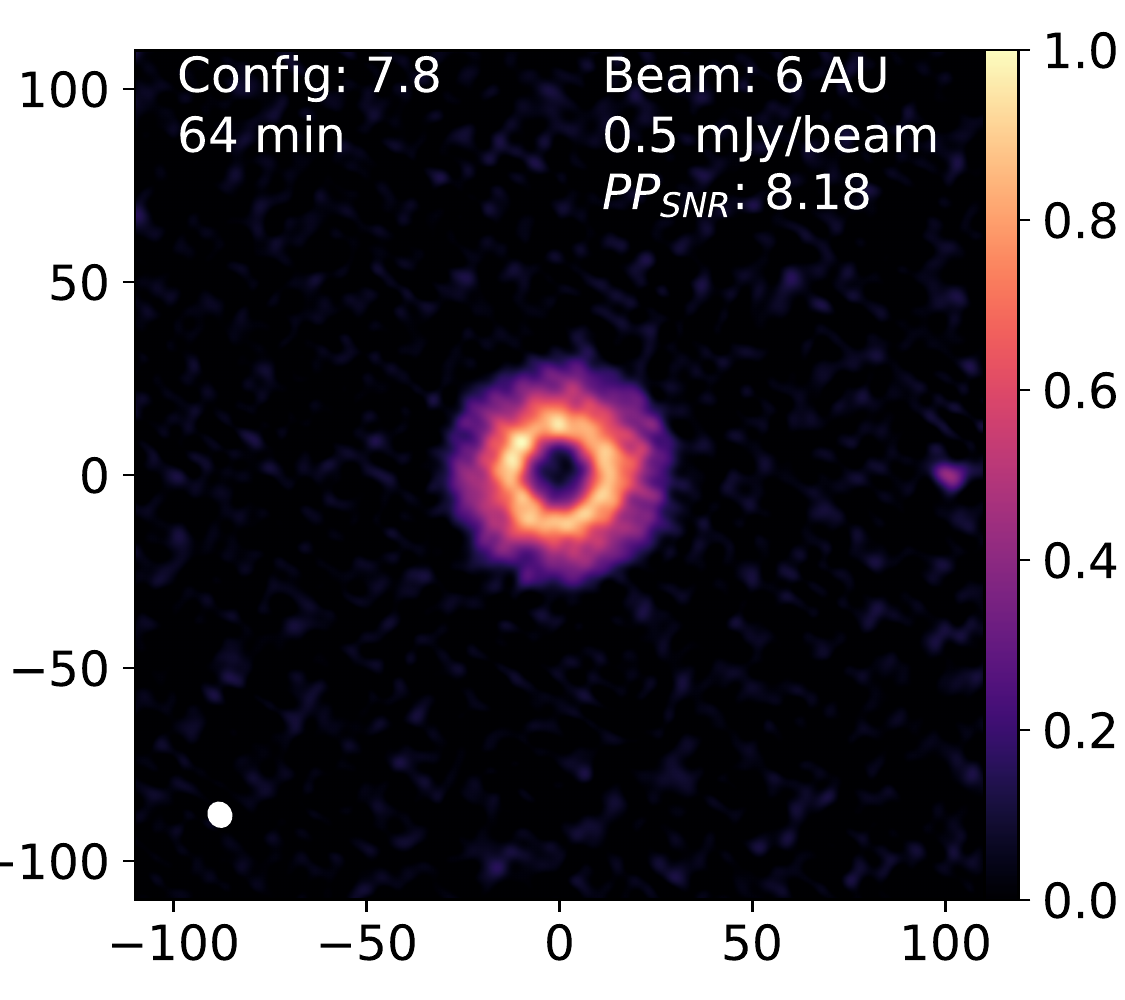}\\
    d) & e) & f) \\
\end{tabular}
\caption{Same as Figure \ref{fig:ALMA_ints} but now for a 3 $\mj$ protoplanet at 100 AU. Protoplanets at wider orbits have lower surface temperatures and are therefore more difficult to detect with ALMA.}
\label{fig:ALMA_ints_100AU}
\end{figure*}

\textbf{Figure \ref{fig:ALMA_ints_100AU} }shows an additional visualisation of the effect of integration time on an observation of a 3 $\mj$ protoplanet at 100 AU, due to its cooler surface temperature of 20 K the protoplanet is more difficult to detect. This figure illustrates the importance of adequate integration time for fainter protoplanets. It is very likely that GI may leave protoplanets stranded at these large radii, the theory requires initial protoplanet formation only takes place beyond 50 AU and protoplanet-protoplanet scattering may increase the orbital radius at early times.

\section{Protoplanet evolution models}
\label{app:models}

\begin{figure*}
\includegraphics[width=1.0\columnwidth]{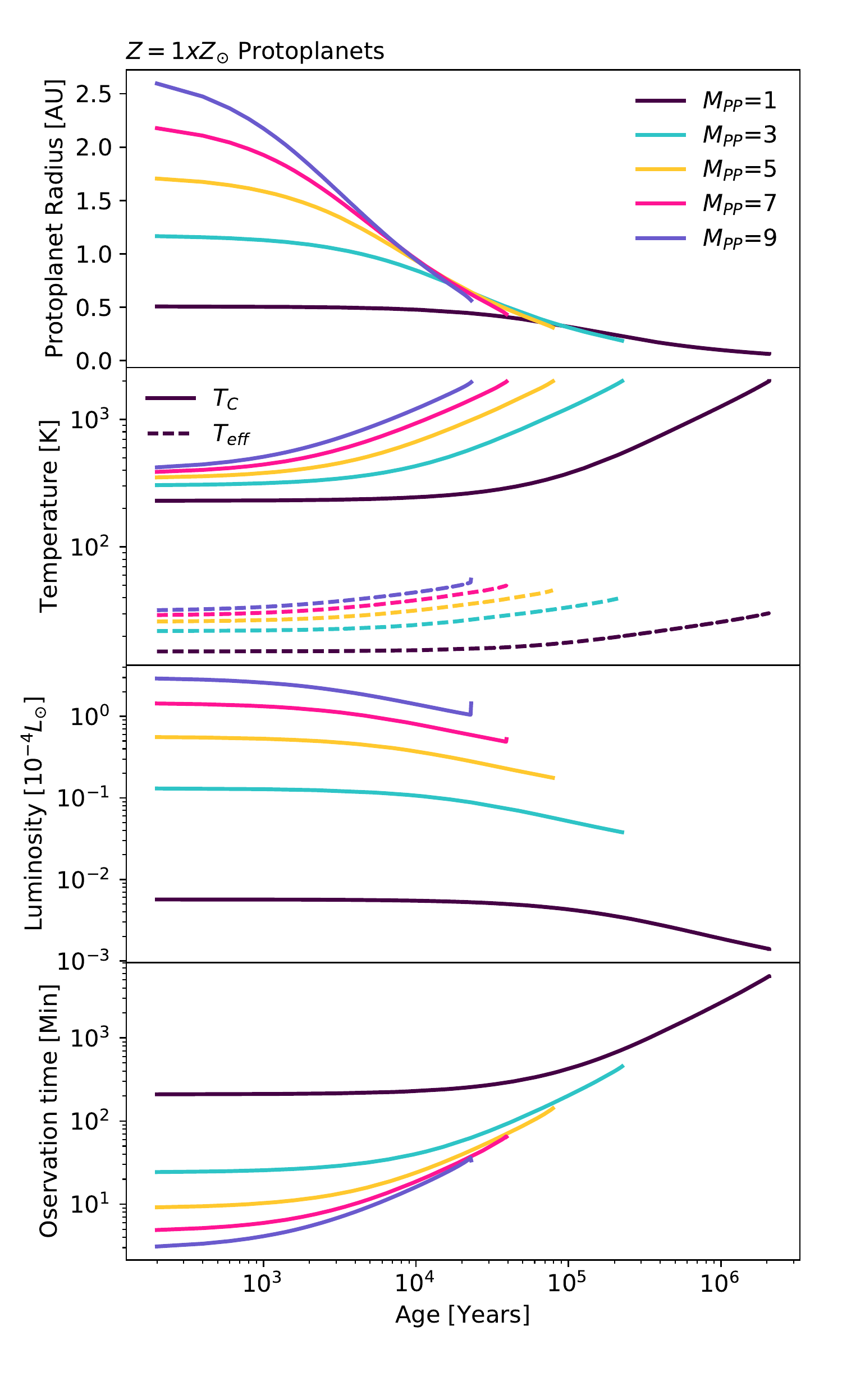}
\includegraphics[width=1.0\columnwidth]{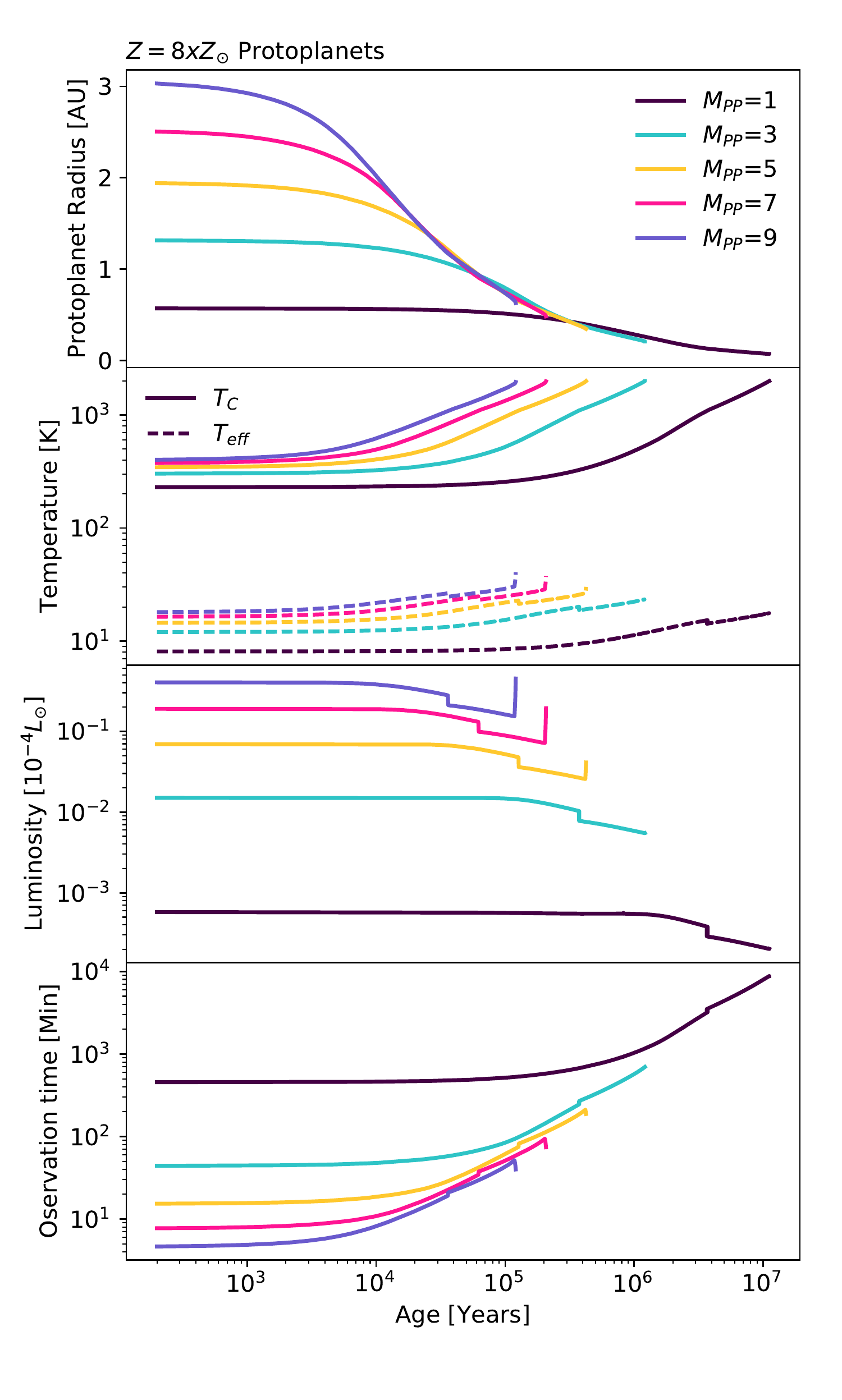}
\caption{Protoplanet evolution models, fully described in \protect\cite{NayakshinFletcher15} and associated papers. Once the central temperature in a protoplanet reaches 2000 K, Hydrogen in its core dissociates and it rapidly collapses to a `Jupiter-like' state. Left: $Z=Z_{\odot} = 0.015$, right: $Z=8 \times Z_{\odot}=0.12$. Observation time is calculated from our radiative transfer calculation, scaled using Equation \ref{eq:obs_time}. Note that high metallicity protoplanets are more opaque, have lower luminosities and therefore survive for longer before collapse.}
\label{fig:PP_models}
\end{figure*}

Figure \ref{fig:PP_models} shows the output of the protoplanet evolution models from \cite{NayakshinFletcher15} and associated papers. In Section \ref{sec:obs} we combined these models with our radiative transfer simulations in order to calculate the necessary observation time across a range of protoplanet ages and masses. The result of Equation \ref{eq:obs_time} is plotted in the bottom panels of these figures. We can see that protoplanets dim considerably over their lifetimes, supporting our conclusion that observations should focus on the youngest systems.

\bsp	
\label{lastpage}
\end{document}